\documentclass[a4paper,10pt,twoside]{cpc-hepnp}

\usepackage{multicol}
\usepackage{graphicx}
\usepackage{booktabs}
\usepackage{amssymb,bm,mathrsfs,bbm,amscd}
\usepackage[tbtags]{amsmath}
\usepackage{lastpage}
\usepackage{color}

\begin{document}

\fancyhead[c]{\small Chinese Physics C~~~Vol. xx, No. x (2018) xxxxxx}
\fancyfoot[C]{\small 1-\thepage}

\footnotetext[0]{Submitted January 2018}

\title{Relativistic compact stars with charged anisotropic matter}

\author{$^1${S. K. Maurya}\email{sunil@unizwa.edu.om}, $^2${Ayan Banerjee}\email{ayan7575@yahoo.co.in}, $^3${Phongpichit Channuie}\email{channuie@gmail.com}
}
\maketitle

\address{$^1$ Department of Mathematical \& Physical Sciences, College of Arts \& Science, University of Nizwa, Nizwa, Sultanate of Oman\\
$^2$Department of Mathematics, Faculty of Applied Sciences, Durban University of Technology, Durban, South Africa\\
$^3$School of Science, Walailak University, Nakhon Si Thammarat, 80160 Thailand}

\begin{abstract}
In this article, we perform a detailed theoretical analysis of new exact solutions with anisotropic fluid distribution of matter for compact objects subject to hydrostatic equilibrium. We present a family solution to the Einstein-Maxwell equations describing a spherically symmetric, static distribution of a fluid with pressure anisotropy. We implement an embedding class one condition to obtain a relation between the metric functions. We generalize the properties of a spherical star with hydrostatic equilibrium using the generalised Tolman-Oppenheimer-Volkoff (TOV) equation. We match the interior solution to an exterior Reissner-Nordstr\"{o}m one, and study the energy conditions, speed of sound, and mass-radius relation of the star. We also show that the obtained solutions are compatible with observational data for the compact object Her X-1. Regarding our results, the physical behaviour of the present model may serve for the modeling of ultra compact objects.
\end{abstract}

\begin{keyword}
Class I spacetime; exact solutions; compact objects; electromagnetic mass models
\end{keyword}

\begin{pacs}
 00.02; 00.04; 90.95
\end{pacs}

\footnotetext[0]{\hspace*{-3mm}\raisebox{0.3ex}{$\scriptstyle\copyright$}2018
Chinese Physical Society and the Institute of High Energy Physics
of the Chinese Academy of Sciences and the Institute
of Modern Physics of the Chinese Academy of Sciences and IOP Publishing Ltd}%

\begin{multicols}{2}

\section{Introduction}
Over the past few decades, the study of relativistic compact stars has received much attention.  In order to model the objects, we examine the solutions of the Einstein equations for statically, spherically symmetric geometry with different physical grounds. These may be refer to, for example, perfect fluid, dust, and anisotropic features. However, there is strong theoretical evidence that suggests that highly dense celestial bodies are not composed purely of perfect fluids. In some cases, the objects are found with different physical phenomena, e.g. anisotropy. The first theoretical attempt at considering the anisotropy effect dates back to around 1922 when Jeans \cite{firstaniso} considered pressure anisotropy in self-gravitating objects  for Newtonian configurations. Moreover, in the relativistic limit, it was considered by Lema\^{i}tre \cite{Lemaitre:1933gd}.  More recently, the anisotropic effect was studied by Ruderman \cite{Ruderma}. He argued that at very high densities, of the order of $10^{15}\,\mathrm{g/cm^3}$, stars may have anisotropic features where the nuclear interaction becomes relativistic. Soon after, in 1974 Bowers and Liang~\cite{Bowers:1974tgi} studied local anisotropic properties for static spherically symmetric and relativistic configurations, which have been extensively populated. For instance, analytical static solutions have been considered by many authors \cite{groups}. For a proper chronological order, we recommend readers to the review article of Herrera and Santos \cite{Herrera:1997plx}, where they have discussed the full details of local anisotropy in self gravitating systems. Much earlier than the work proposed by Mak and Harko \cite{Mak}, a plethora of exact solutions for an anisotropic fluid sphere have been given in the literature \cite{Cosenza1981,Herrera1984,Herrera1985-2,Herrera1985-1,Esculpi1986,Herrera:2001vg,Herrera:2002bm}. Notice that the generalized relativistic analogue for anisotropic stars was considered by several authors, e.g. Refs.~\cite{Sharma,FR1,Banerjee,SG1,Matondo} and references therein. It is found that the solutions have been further developed based on the same method \cite{Ivanov,M1,Jose,B1,FR2,SH}. More interestingly, the algorithm for all static anisotropic solutions and charged anisotropic solutions of Einstein's equation in the spherically symmetric case can be nicely obtained by a general method introduced by Refs. \cite{Herrera:2007kz} and \cite{Maurya2017aa}.

The theoretical motivation for the concept of an embedding general $n$-dimensional manifold in higher-dimensional spacetime is now an interesting problem that plays an essential role in studying the structure of stars.
In particular, the primary motivation for the embedding problem emerged from the work done
by Schl\"{a}fli \cite{Schlafli}, soon after the publication by Riemann. He considered the
problem of how to locally embed such manifolds in Euclidean space $E_N$ with $N = n(n + 1)/2$.
A central aspect of this embedding is not only to search for new solutions, but also to look for higher-dimensional generalizations of our known four-dimensional solutions. A few examples in the generalization,
such as Whitney's embedding theorem \cite{Whitney}, claim that any $n$-dimensional manifold
can be embedded in  $R^n$ (see the Janet-Cartan theorem \cite{Janet,Cartan}) insisting that any real analytic $n$-dimensional
Riemannian manifold can be locally embedded in $n(n+1)/2$-dimensional
Euclidean spaces; and Nash's embedding theorem, which shows that Riemannian manifolds can always
be regarded as Riemannian submanifolds, and be isometrically embedded in some Euclidean spaces.
More recently, some other embedding theorems have been extented to study the
braneworld consequences in general relativity. The braneworld models, based on the
assumption that four dimensional space-time is a 3-brane (domain wall), are embedded in a
5-dimensional Einstein space \cite{Randall,Maartens}. This framework has been successfully used in
the context of astrophysical research as well as in cosmological paradigms.

Moreover, it is well known that an $n$-dimensional manifold $V_n$ can be embedded in $m =n(n + 1)/2$-dimensional pseudo-Euclidean space, which is called the embedding class of $\mathrm{V}_n$. Inspired by this classification,
the embedding of 4-dimensional spacetime into 5-dimensional flat space-time has been achieved by using the spherical coordinates transformation, which is known as the embedding class one condition
satisfying Karmakar's condition \cite{Karmarkar}. In this case, the metric functions
are connected to each other and we can obtain exact solutions of Einstein's equation reducing
to a single-generating function. This method is adopted by many authors, in the simple
model of spherically symmetric stars with anisotropic pressure
\cite{Maurya(2016),Singh,Bhar}. However, the discovery of several peculiar compact stellar objects, such as the
X-ray pulsar Her X-1, X-ray burster 4U 1820-30, X-ray sources 4U 1728-34,
PSR 0943+10, millisecond pulsar SAX J 1808.4-3658, and RX J185635-3754, motivate us to investigate the model and compare the predictions to the state-of-the-art stellar models. More recently, some models explaining compact objects have often been formulated on various physical grounds including charged and uncharged with different equations of state (EOS) \cite{sgov,npg,M2,M3,M4,M5,K1,K2,K3,pandey1}. The aim of the present paper is to obtain a new class of solutions for a charged fluid sphere using the embedding class one conditions. It is found that these models coincide with the compact stellar object Her X-1, presented in Gangopadhyay \emph{et al} \cite{Gangopadhyay}.

This paper is structured as follows. In Section 2, we propose the class one condition and
the metric functions related  in a closed form by using the class
one condition. In Section 3, we set up the relevant field equations and their solutions
for a charged fluid distribution. In Section 4, we derive an exact solution
of the Einstein-Maxwell equations by assuming the  simplest form of an
electric field $E$. We then present physical analysis of stellar models by comparing the results with the observational
stellar mass for Her X-1 and discuss the stability of the charged stars by
employing modified TOV equations. We finally summarize the results in the last section.

\section{Embedding class one condition and  Einstein-Maxwell field equations for charged anisotropic matter distribution}
As suggested in Ref. \cite{Maurya:2017sjw}, the line element that describes a static, spherically symmetric matter distribution is in general written in terms of the Schwarzschild $x^i = (r, \theta, \phi, t)$ as follows:
\begin{equation}
ds^2 =  e^{\nu(r)} dt^2- e^{\lambda(r)} dr^2 - r^2(d\theta^2 + \sin^2\theta d\phi^2),\label{metric1}
\end{equation}
where the unknown functions $\nu(r)$ and $\lambda(r)$ have to be determined by solving the Einstein field equations. The implementation of the class one condition of the above metric can be successfully achieved by considering a five-dimensional flat line element as:
\begin{eqnarray} ds^{2}=-\left(dx^1\right)^2-\left(dx^2\right)^2-\left(dx^3\right)^2-\left(dx^4\right)^2+\left(dx^5\right)^2,\label{e1}
\end{eqnarray}
where we have transformed the coordinate system via the following rescaling: $x^1=r\,\sin\theta\,\cos\phi$, \, $x^2=r\,\sin\theta\,\sin\phi$, \,$x^3=r\,\cos\theta$,\, $x^4=\sqrt{K}\,e^{\frac{\nu}{2}}\,\cosh{\frac{t}{\sqrt{K}}}$,\, $x^5=\sqrt{K}\,e^{\frac{\nu}{2}}\,\sinh{\frac{t}{\sqrt{K}}}$, with $K$ being a positive constant. With the differential forms of the given components, we come up with the following expression for the above metric:
\begin{equation}
ds^{2}=e^{\nu(r)}dt^{2}-\left(1+\frac{K e^{\nu}}{4} {\nu'}^2 \right) dr^{2}-r^{2}\left(d\theta^{2}+\sin^{2}\theta d\phi^{2} \right),\label{e2}
\end{equation}
which features a four-dimensional and spherically symmetric line element. This metric describes a  5D pseudo-Euclidean space embedded in a (3+1)-dimensional spacetime. By equating the line elements given in Eq.~(\ref{metric1}) and Eq.~(\ref{e2}), one obtains the following relation:
\begin{equation}
e^{\lambda}=\left(\,1+\frac{K\,e^{\nu}}{4}\,{\nu'}^2\,\right).\label{eq4}
\end{equation}
Equation~(\ref{eq4}) provides the embedding class condition (for more details, we refer the reader  to Refs. \cite{Karmarkar,M2,Maurya(2016)}). For completeness we give a short recap by writing the Einstein-Maxwell field equations on a four dimensional space-time manifold:
\begin{equation}
 {R^\mu}_{\gamma} - \frac{1}{2} R\, {g^\mu}_{\gamma}= 8\,\pi ({T^\mu}_{\gamma}+ {E^\mu}_{\gamma}).\label{field}
\end{equation}
In the following, we will use the geometrized units $G = c = 1$ with $G$ and $c$ being the Newtonian gravitational constant and speed of light in vacuum, respectively. In our present consideration, we include the energy-momentum tensor for the Maxwell fields and the complete form of anisotropic charged fluid matter is given by:
\begin{eqnarray}
{T}^{\mu}_{\gamma} &=& \left[(\rho + p_t)u^\mu\,u_{\gamma} - p_t{\delta^\mu}_{\gamma} + (p_r - p_t) \chi^\mu \chi_{\gamma}\right],\,\\
{E^\mu}_{\gamma} &=& \frac{1}{4\pi}\left[- F^{\mu\,m}F_{\gamma m} + \frac{1}{4}{\delta^\mu}_{\gamma}\,F^{mn}F_{mn}\right],\,
\end{eqnarray}
where $u^{\mu}$ is the four-velocity of the fluid, given by $u^{\mu}=e^{\nu(r)/2}\delta^{\mu}_{4}$, and $\chi^{\mu}$ is the unit space-like vector in a radial direction, given by $\chi^{\mu}=e^{\lambda(r)/2}\delta^{\mu}_{1}$. Here $\rho,\,p_{r}$ and $p_{t}$ are the energy density, the radial pressure and the tangential pressure, respectively. As usual, the 4-velocity and the radial unit
vector satisfy the following the conditions:
\begin{equation}
u^{\gamma}\,u_{\gamma} =-1,~~~~ \chi^{\gamma}\,\chi_{\gamma}=1 ,~~~~ \chi^{\gamma}\,u_{\gamma}=0.
\end{equation}
However, the other components (radial and spherical) of the 4-vector are absent. The components for ${T^\mu}_{\gamma}$ and ${E^\mu}_{\gamma}$ are defined as follows:
\begin{equation}
 {T^1}_1=-p_r,\, {T^2}_2={T^3}_3=-p_t,\, {T^4}_4=\rho,\,\label{Eq9}
 \end{equation}
 \begin{equation}
 {E^1}_1=-{E^2}_2=-{E^3}_3={E^4}_4=\frac{1}{8\,\pi}\,e^{\nu+\lambda}\,F^{14}\,F^{41}.\label{Elec}
\end{equation}
Now, the Maxwell field equations give the following relationship:
\begin{equation}
\left[\sqrt{-g}F^{\mu \varphi}\right]_{,\varphi}=4\pi j^{\mu}\sqrt{-g},
 \end{equation}
where $j^{\mu}$ is the four-current density. For a particular choice of the electromagnetic field, the only non-vanishing components of the electromagnetic field tensor are $F^{01}$ and $F^{10}$, and are related by $F^{01} = - F^{10}$. The Maxwell field-strength tensor, $F^{\mu \gamma}$, can be written in terms of a four-potential $A^{\mu}$: $F_{\mu \gamma}$ = $A_{\mu, \gamma}$ - $A_{\gamma, \mu}$. So, for a non-vanishing field tensor, only the existing potential is $A_0$ = $\phi$. In addition, the potential considered
 has a spherical symmetry, i.e., $\phi$= $\phi(r)$. Now using  Eq.~(\ref{Elec}), one can obtain the electric field as
\begin{equation}
F^{01}(r)= E(r)= e^{-(\nu+ \lambda)/2}r^{-2}\int^{r}_{0}{4\pi j^{0}e^{(\nu+ \lambda)/2}}dr^{\prime}. \label{E12}
\end{equation}
Let us consider the total charge, $q(r)$, inside a sphere of radial coordinate $r$:
\begin{equation}
q(r)= \int^{r}_{0}{4\pi j^{0} r^{\prime 2} e^{(\nu+ \lambda)/2}}dr^{\prime}, \label{E13}
\end{equation}
From the above expression, we can relate the electric field and the total charge by using Eqs.~(\ref{E12}) and (\ref{E13}) and find
\begin{equation}
E(r)= e^{-(\nu+ \lambda)/2}r^{-2} q(r). \label{Er}
\end{equation}
Due to the effects of the electric field and pressure anisotropy, the Einstein-Maxwell field equations with the metric (\ref{metric1}) provide the following expressions
\begin{equation}
8\,\pi\, \rho (r) =\frac{K\,e^{\nu } \nu '\,[ 4\,r\,(2\nu '' +\nu ^{\prime2} )+\nu '\,(4+K\nu ^{\prime2} e^{\nu } )]}{r^2\,(4+K\nu ^{\prime2} e^{\nu } )^2}
- \frac{q^{2} }{r^{4} }, \label{rho}
\end{equation}
\begin{eqnarray}
\hspace{-3cm}
8\,\pi\, p_r(r) =  \frac{\nu ' \,(4r-K\nu 'e^{\nu}) }{r^{2} (4+K\nu ^{\prime2} e^{\nu } )} +\frac{q^{2} }{r^{4} },  \label{pr}
\end{eqnarray}
\begin{equation}
8\,\pi\, p_t(r)=  \left[ \frac{2 \nu ' (4+K\nu
^{\prime2} e^{\nu } )-2 (K\nu ' e^{\nu } -2r) (2\nu '' +\nu ^{\prime2} ) }{r\,(4+K\nu
^{\prime2} e^{\nu } )^2}\right] - \frac{q^{2} }{r^{4} },\label{pt}
\end{equation}

where primes denote derivatives with respect to the radial coordinate $r$. Following the method given in Ref.~\cite{Maurya:2017sjw}, the solution for the potential metric $e^{-\lambda(r)}$ is given by
\begin{equation}
e^{-\lambda(r)} = 1 -\frac{2m(r)}{r} + \frac{q^2}{r^2},\label{e4}
\end{equation}
where the parameters $q$ and $m$ represent the charge and mass within the radius $r$, respectively.
In the present case, considering Eqs.~(\ref{pr}) and (\ref{pt}), we can obtain the following form
\begin{eqnarray}
8\,\pi\,\Delta +\frac{2q^{2} }{r^{4} }=
\left[ \frac{\nu '\,(4+K\nu ^{\prime2} e^{\nu })-2\,r\,(2\nu '' +\nu
^{\prime2} ) }{r^2\, \left({K\nu ' e^{\nu }-2\,r} \right)^{-1}\, (4+K\nu ^{\prime2} e^{\nu } )^2} \right] , \label{eq21}
\end{eqnarray}
where $\Delta$ = $p_t - p_r$ denotes the anisotropy factor, which measures the pressure anisotropy of the fluid. It is important to note that $\Delta = 0$ at the origin, $r = 0$, corresponds to the particular case of an isotropic pressure. However, in the absence of both anisotropy and charge throughout the interior of the star, the right counterpart of Eq.~(\ref{eq21}) gives only two kinds of perfect fluid, namely the Schwarzschild interior solution \cite{Schwarzschild} and the Kohler-Chao solution \cite{Kohler}.
Moreover the factor $\Delta/r$ represents a force due to the anisotropic nature of the fluid. The anisotropy will be
repulsive when $p_t > p_r$, and attractive when $p_t < p_r$.

It is clear from Eqs.~(\ref{Er}-\ref{eq21}) that we have six unknown functions, namely, $\rho(r)$, $p_r(r)$, $p_t(r)$, $\nu(r)$, $\lambda(r)$  and $q^2(r)$, with three independent equations.  In attempting to find exact solutions, there are two different approaches we can take: (i) one can choose a specific mass function $m(r)$, by specifying an equation of state (EOS) of the form $p = p(\rho)$; or (ii) one can establish a relation between gravitational potentials $\nu$ and $\lambda$ based upon physical grounds in this case of a charged fluid. In this work, in order to obtain a unique solution, we consider embedding class one conditions that
relate a close connection between two metric functions, as discussed in this section, for an electrically charged fluid sphere. Some recent investigations have been considered in Ref. \cite{P1,M6,M7}, where the authors studied compact objects with anisotropic pressure via embedding.

\section{Generalized charged anisotropic solution for compact star}\label{f3}

In this present work, we will examine the solutions for a charged fluid model by considering a
single generic function $\lambda(r)$. The ansatz has a geometric
interpretation from the physical point of view and was previously used to obtain
solutions for compact stellar objects \cite{M8,Abbas,Zubair}.
However, we are interested in the solution that can be used to model the compact object Her X-1. Following Ref.~\cite{Maurya:2017sjw}, the system we are going to study here is characterized by the fact that $\lambda(0) = \textit{finite constant}$ for a charged compact star and energy density $\rho(r)$, radial pressure $p_r(r)$, and tangential pressure $p_t(r)$ should be finite at the origin. For physical grounds, we assume that the electric field at the centre is zero and $m(0)=0$ with $m^{\prime}(r) > 0$.
Let us now suppose the generic function $\lambda = \lambda(r)$ with the following expression Ref.~\cite{Maurya:2017sjw}:
\begin{eqnarray} \label{n10}
\hspace{-1cm}\,\lambda(r)=\ln\Bigl(1+a\,r^2\,(1+b\,r^2)^{2 n}\Bigl).
\end{eqnarray}
In this particular approach, we keep in mind that $\lambda^{\prime}(r)=0$ and $\lambda^{\prime\prime}(r) > 0$.
Now, substituting the value of $\lambda$ into Eq.~(\ref{eq4}), we obtain
\begin{equation}
\nu(r)= 2\,\ln[A+B\,(1+b\,r^2)^{1+n}],\label{eq10}
\end{equation}
where $A$ is an integration constant and $B
=\sqrt{a}\left(4\,b\,\sqrt{K}\,(n+1)\right)$. Notably, we find
\begin{eqnarray}
 \nu(0)&=&2\,\ln(A+B),\\
\nu' (r) &=&\frac{4\,r\,b\,B\,(n+1)\,(1+br^2)^n}{[A+B\,(1+br^2)^{n+1}]},
\end{eqnarray}
and
\end{multicols}

\begin{eqnarray}
\nu''(r) = \frac{4\,(n+1)\,b\,B\,(1+ b\,r^2)^{n-1}[A\,(1 + b\,r^2 + 2\,n\,b\,r^2)- B\,(1 + b\,r^2)^n (b^2 r^4-1)]}{[A + B\,(1 + b\,r^2)^{1 + n}]^2}.
\end{eqnarray}

\begin{multicols}{2}
The above decompositions clearly imply that $\nu(0)> 0$, $\nu'(0) = 0 $ and $\nu''(0) = \frac{4\,(n+1)\,b\,B}{(A+B)} >0$, which provides $(A+B) > 0$. The motivation behind this particular choice of metric function (\ref{n10}) is not new in stellar modelling.

Similar forms for the metric functions have previously been
considered by Singh \textit{et al.} \cite{K2} and Maurya \textit{et al.} \cite{M6} for positive and negative powers of $n$
for anisotropic matter distribution using the embedding class one condition.

Note that the parameter $n$ plays an important role in determining the structure and stability of the compact object. Here, we will consider the following three cases: (i) when $\mathrm{n} >0$ and $b > 0$;
(ii) when $n>0$ and $b > 0$; and (iii)  when $-1\le n < 0$ and $b\in \Re$.
The choices of the range of free parameters are reasonable in the sense that they
admit physically viable models. In the relativistic regime,
we describe here stellar compact objects such as Her X-1, which provide the choice of parameters
for a charged anisotropic fluid sphere. Furthermore, we assume the general form of an electric field intensity $E$ within
the radius $r$ as
\begin{eqnarray}
\hspace{-3.2cm}
E^2=\frac{q^2}{r^4}=q_0\,a^2\,r^4\,(1+b\,r^2)^{2\,n},
\end{eqnarray}

where $q_0$ is a positive constant. The expression is invariant under the transformation $E \rightarrow - E$,
and only positive values of $E$ are chosen. It is clear from the above expression that the electric field $ E$
vanishes at the centre, i.e., $r=0$. As already mentioned, our model consists of a charged relativistic object, with
anisotropic matter distribution. Making use of this assumption, we obtain the mass function using Eq.~(\ref{e4}), which yields
\begin{eqnarray}
\hspace{-0.1cm}
m(r)=\frac{r\,(1+b\,r^2)^{2\,n}}{2}\left[\frac{a\,r^2\,}{[1+a\,r^2\,(1+b\,r^2)^{2\,n}]}+ q_0\,a^2\,r^6\,\right]
\end{eqnarray}

Note that $m(0) = 0$ at the centre. However, both $q (r)$ and $m (r)$ are positive when
$r > 0$ for all the above three cases of $n$. This indicates that $q(r)$ and $m(r)$ are increasing monotonically away from the centre and attain regular minima at $r = 0$. Utilizing the expression in Eq.~(\ref{n10}), we can now determine the Einstein-Maxwell system of Eqs.~(\ref{rho})-(\ref{pt}) for charged anisotropic matter with the line element (\ref{metric1}), resulting in the following relations:
\end{multicols}

\begin{eqnarray}
\hspace{-2.cm}
\rho (r) &=& -\frac{q_0\,a^2\,x^2\,\Psi^{2\,n}}{8\,\pi}+\frac{1}{8\,\pi}\,\left[ \frac{a\,\Psi^{2 n}}{(1 + a\, x\,\Psi^{2 n})}+\frac{2\,a\,\Psi^{2 n-1}\,(\Psi+2\,n\,x)}{(1 + a\, x\,\Psi^{2 n})^2}\,\right], \\
\hspace{-1.2cm}
p_r (r) &=& \frac{q_0\,a^2\,x^2\,\Psi^{2\,n}}{8\,\pi}+ \frac{1}{8\,\pi}\,\left[\frac{4\, b\, B\,(1 + n)\, \Psi^n}
{(1 + a\, x\, \Psi^{2 n})\, (A + B\, \Psi^{1 + n})}- \frac{a\,\Psi^{2 n}}{(1 + a\, x\,\Psi^{2 n})}\,\right], \\
\hspace{-0.2cm}
p_t (r)&=& -\frac{q_0\,a^2\,x^2\,\Psi^{2\,n}}{8\,\pi}+\frac{\Psi^{n-1}}{8\,\pi} \left[\frac{4\, b\, B\,(1 + n)\,(\Psi+b\,n\,x)\,(1 + a\, x\, \Psi^{2 n})-a\,\Phi(x)}{(1 + a\, x\, \Psi^{2 n})^2\, (A + B\, \Psi^{1 + n})}\,\right],
\end{eqnarray}
where $\Psi=(1+b\,x)^n$ with $a, b$, and $n$ being arbitrary constants.

\begin{figure}[h]
\centering
\includegraphics[width=8cm]{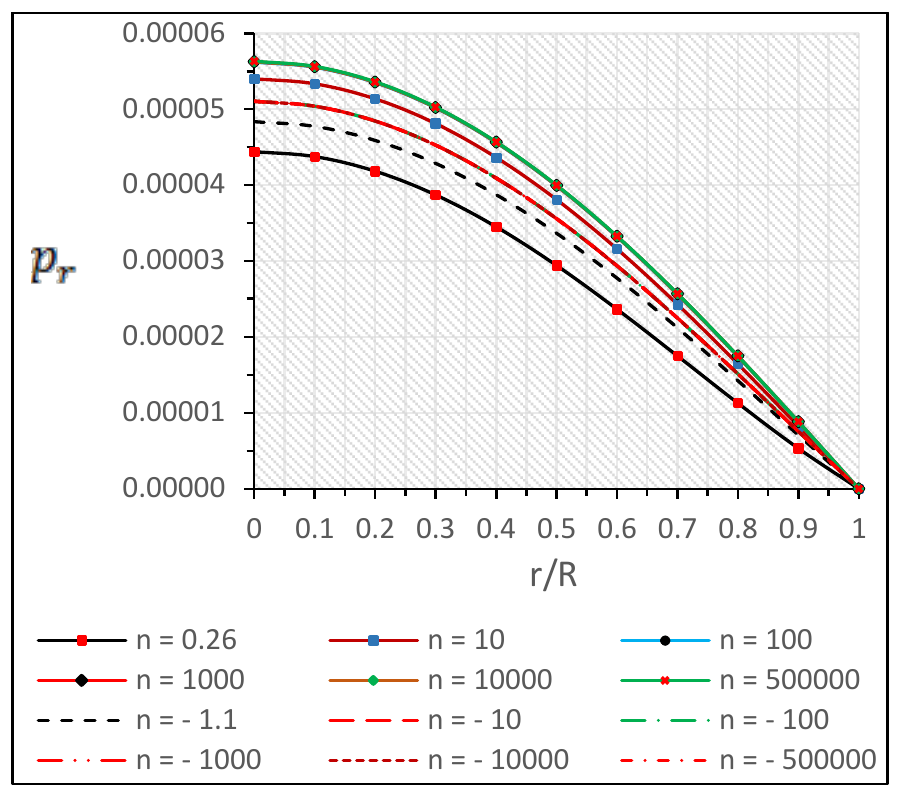} \includegraphics[width=8cm]{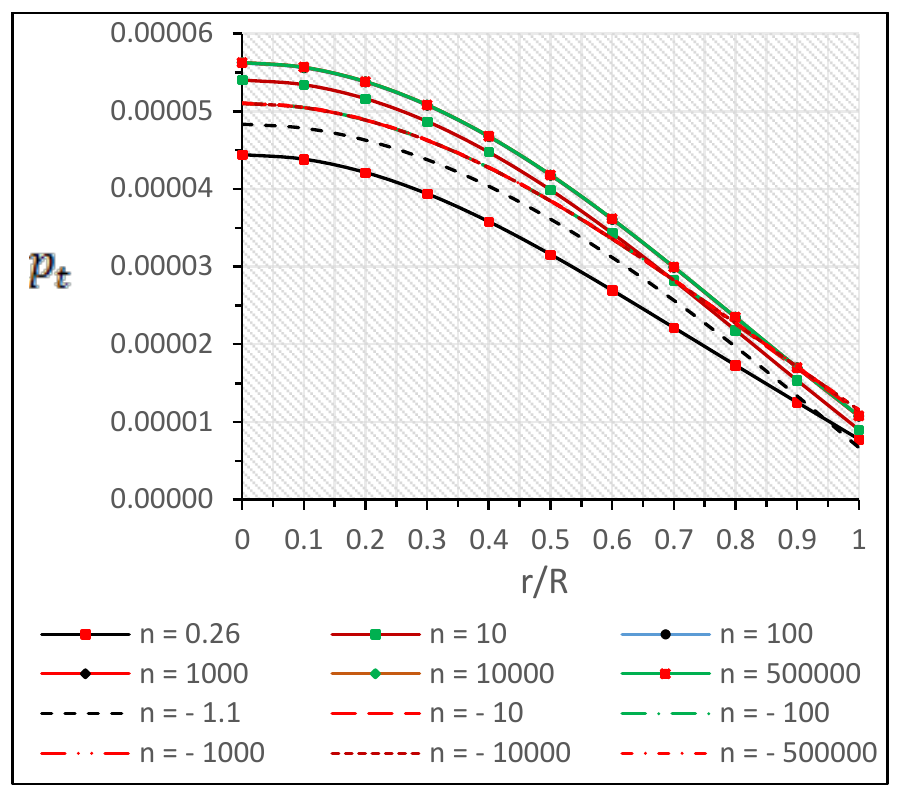}
\caption{Behavior of radial pressure ($p_r$) and tangential pressure ($p_t$) verses fractional radius $r/R$ for Her X-1. The numerical values of the constants are given in Tables 1 and 2.} \label{f1}
\end{figure}
\begin{figure}[h]
\centering
\includegraphics[width=8cm]{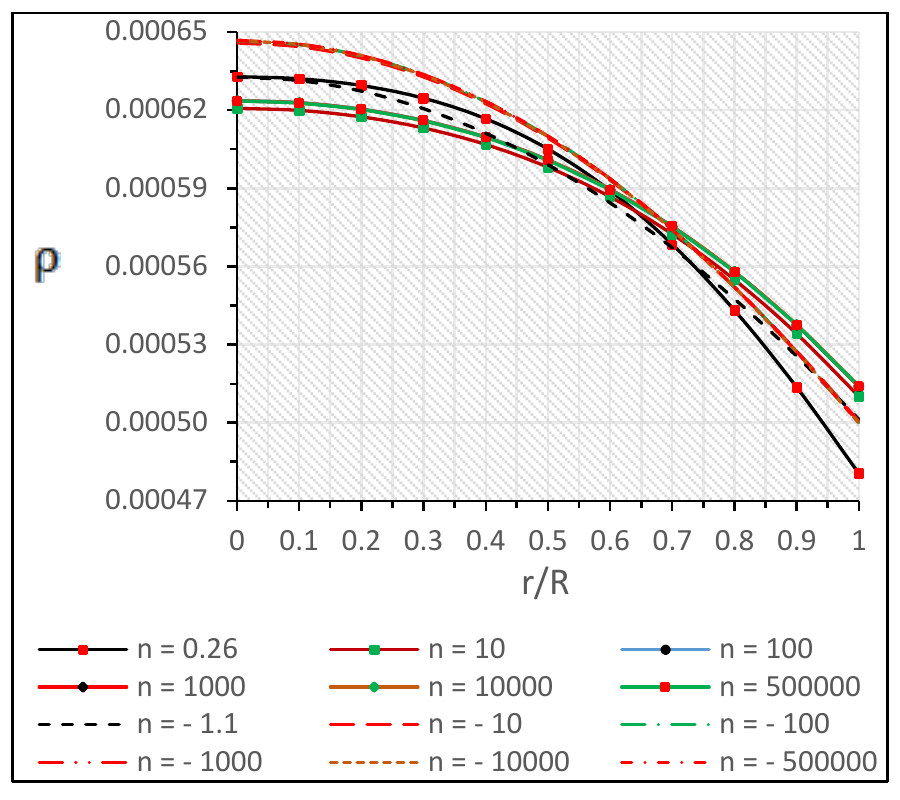} \includegraphics[width=8cm]{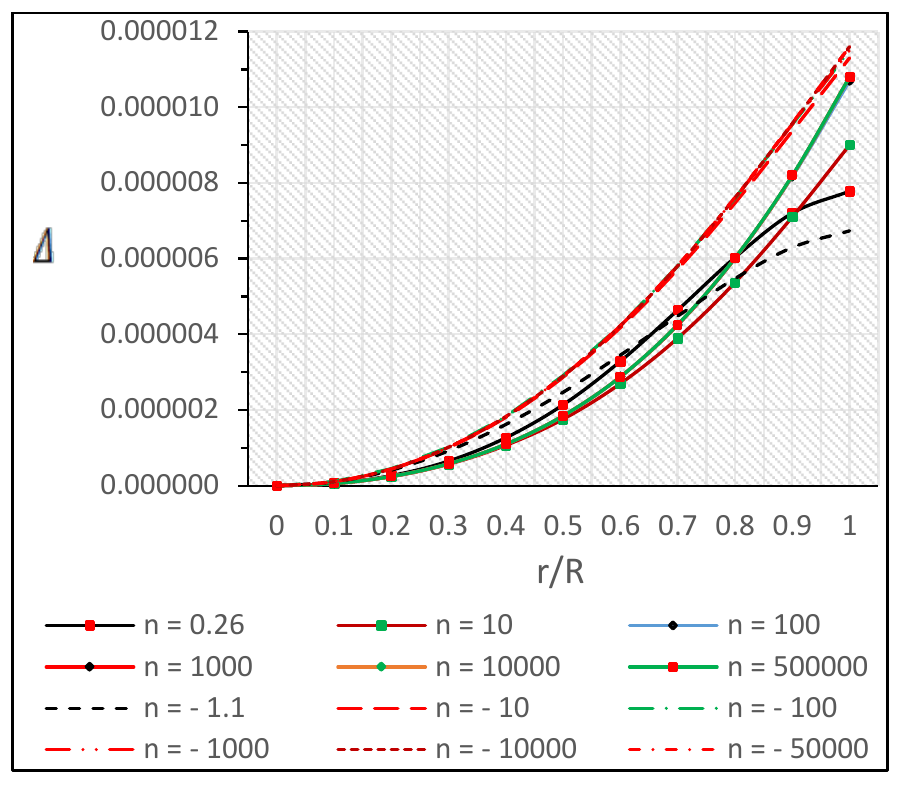}
\caption{Behavior of energy density ($\rho$) and anisotropy factor ($\Delta$) verses fractional radius $r/R$ for Her X-1. The numerical values of the constants are given in Tables 1 and 2.}  \label{f2}
\end{figure}

\begin{multicols}{2}
Let us focus on the effect of the pressure anisotropy, a term $ \Delta = p_t - p_r$, which in terms of the expression in Eq. (\ref{n10}) is recast as
\begin{equation}
\Delta = -\frac{q_0\,a^2\,x^2\,\Psi^{2\,n}}{4\,\pi}+ \frac{x\,\Psi^{n-1}\, (-2\, b\, n + a\,\Psi^{2 n} +
   a\,b\,x\,\Psi^{2 n})\,\Delta_1(r)}{8\,\pi\,(1 + a\, x\,\Psi^{2 n})^2\,(A + B\, \Psi^{1 + n})}. \label{DL1}
\end{equation}
We further consider the following terms:
\begin{eqnarray}
\Phi(x)=\Psi^{n}\,(\Psi+2\,n\,b\,x)\,[A + B\,\Psi^n\,(1 + 3\,b\,x\,+ 2\, n\,b\,x)],\\
\Delta_1(x)=[a\,\Psi^n\,(A + B\,\Psi^n)-b\,B\,(2 + 2\,n - a\, x\,\Psi^{2 n})],
\end{eqnarray}
which feature forces that arise due to the anisotropic nature of the fluid, and maintain the stability and equilibrium configuration of the  stellar structure. Also, from Eq.~(\ref{DL1}) we may observe that the anisotropy $\Delta$ vanishes throughout inside the star if and only if $a=0$, which implies that all pressures, density and mass will become zero and the metric turns out to be flat.

Let us emphasize again the behavior of the energy density and pressures for the
X-ray pulsar Her X-1 with graphical representation. The variation of pressure
and density with radial distance, drawn using Eqs. (\ref{rho}-\ref{pr}), are shown in Figs.~\ref{f1} and \ref{f2}, respectively.
We have used the data set for model parameters given in Table\,\ref{Table1-0}\,\&\,\ref{Table1-1}.
In this case one can see that energy density is positive and the
radial pressure $p_r$ vanishes at the boundary of the star for different values of $n$ within the above mentioned specific range .
\end{multicols}

\begin{table}
\centering \caption{Numerical values of physical parameters when $n$ is positive and mass $M =0.85\left(M_\odot\right)$, radius $R=8.10~(km)$ ~\cite{Gangopadhyay}.} \label{Table1-0}

{\begin{tabular}{@{}ccccccc@{}} \hline

 $n$  & $a$   &   $b$   &  $q_{0}$   &  $A$  &   $B$  &    $K$  \\ \hline

$0.26$ & 0.005302 & 0.008 & 0.0007396 & 0.6317681 & 0.11961 & 2.2796 $\times10^{2}$  \\ \hline

$10$ & 0.005200 & 0.000201 & 0.0002410 & 0.1929809 & 0.553565 &   2.1695 $\times10^{2}$ \\ \hline

$10$ & 0.005200 & 0.000201 & 0.0002410 & 0.1929809 & 0.553565 &   2.1695 $\times10^{2}$  \\ \hline

$100$ &0.005225 & 0.00002016 & 0.00008999 & 0.1377625 & 0.606989 & 2.1378$\times10^{2}$  \\ \hline

$1000$ & 0.0052245 & 0.000002016 & 0.00008598 &0.1322219 & 0.612510 & 2.1372 $\times10^{2}$  \\ \hline

$500000$ & 0.0052246 & 0.000000004032 & 0.00008444 & 0.1315549 & 0.613166 &  2.1370 $\times10^{2}$ \\ \hline

\end{tabular}}
\end{table}


\begin{table}
\centering \caption{Numerical values of physical parameters when $n$ is negative and mass $\mathrm{M}=0.85\left(M_\odot\right)$, radius $R=8.10~( km)$ ~\cite{Gangopadhyay}.} \label{Table1-1}

{\begin{tabular}{@{}ccccccc@{}}  \hline
 $n$  & $a$   &   $b$   &  $q_{0}$   &  $A$  &   $B$  &    $K$  \\ \hline

$-1.1$ & 0.0053 & -0.00152 & 0.00049173 & -7.278674 & 8.027761   & 2.22474$\times10^{2}$   \\ \hline

$-10$ & 0.00541 & -0.0001683 & 0.000241 & -0.078002 & 0.823826  & 2.17146 $\times10^{2}$ \\ \hline

$-100$ &0.0054175& -0.00001683 &0.000234 & -0.0039761 & 0.749693 &2.17007 $\times10^{2}$  \\ \hline

$-1000$ & 0.005418 & -0.000001683 & 0.0002332 & 0.00272656 & 0.7429889 & 2.16998$\times10^{2}$  \\ \hline

$-10000$ & 0.0054185 & -0.0000001683 &0.00023181 & 0.0032851 & 0.7424096  & 2.16965 $\times10^{8}$  \\ \hline

$-500000$ & 0.0054188 & -0.000000003366 & 0.000230444 & 0.00327454 & 0.7424053  & 2.16937$\times10^{10}$  \\ \hline

\end{tabular}}
\end{table}
\begin{multicols}{2}

\section{Physical features and comparative study of the physical parameters
for compact star model}
In this section, we will investigate the properties of high density stars
based on the obtained solutions, with the internal structure of the stars satisfying
some general physical requirements, i.e., energy conditions, hydrostatic stability, the speed of sound and the maximum mass of compact objects, by analytical expression as well as graphical representations.
In this manner, we model a compact star composed of anisotropic matter, which
coincides with compact stars like Her X-1.

\subsection{Boundary conditions}
To study a static, spherically symmetric charged star, we match the interior spacetime to an exterior vacuum Reissner-Nordstr\"{o}m (RN) spacetime at the junction surface with radius $r = R$, i.e. matching of first fundamental form (continuity of metric potentials) and second fundamental form (continuity of $\frac{\partial g_{tt}}{\partial r}$) at the surface of the star with radius $r = R$ (Darmois-Israel condition). The RN metric is given by
\begin{eqnarray}
    \label{eq15}
ds^{2} =\left(1-\frac{2M}{r}+\frac{q^2}{r^2}\right)\, dt^{2} -\left(1-\frac{2M}{r}+\frac{q^2}{r^2}
\right)^{-1} dr^{2} \nonumber \\
-r^{2} (d\theta ^{2} +\sin ^{2} \theta \, d\phi ^{2} ),
\end{eqnarray}
where $M$ denotes the total mass of the compact star. We consider standard matter with a spherically symmetric anisotropic fluid where the radial pressure $p_{r}$ must be finite and positive inside the star, and vanishes at the boundary $r = R$ of the star (which is known as the second fundamental form)~\cite{Misner}.
Thus, the second fundamental form i.e. the radial pressure $p_r(R) = 0$, now gives
\begin{equation}
\frac{A}{B}=\frac{a (1+b\,R^2)^{2n} [1-a q_0 R^4+a^2 q_0 R^4\,(1+b R^2)^{2n}]-F_1(R)}
{a\,(1+b\,R^2)^{n}\,[-1+a\,q_0\,R^4+a^2\,q_0\,R^6\,(1+b\,R^2)^{2n}]},
\end{equation}
where $F_1(R)=b\,(4+4\,n-a\,R^2\,(1+b\,R^2)^{2n}+a^2\,q_0\,R^6\,(1+b\,R^2)^{2n}+a^3\,q_0\,R^8\,(1+b\,R^2)^{2n}]$. We can consider the continuity of the first fundamental form
 across a surface at $r = R$, implying that $g_{tt}^{+}$ = $g_{tt}^{-}$ and using the condition
 $e^{-\lambda(R)}=e^{\nu(R)}$ to yield
\begin{equation}
B=\frac{1}{\left[\frac{A}{B}+(1+b\,R^2)^{1+n}\,\sqrt{1+a\,R^2\,(1+b\,R^2)^{2\,n}}\right]}.
\end{equation}
Therefore, one can easily get the total mass of the star for the static case when $e^{-\lambda(R)}=1-\frac{2M}{R}+\frac{Q^2}{R^2}$ as
\begin{equation}
M(R)=\frac{R}{2}\left[\frac{a\,R^2\,+ q_0\,a^2\,R^6\,(1+b\,R^2)^{2\,n}\,[1+a\,R^2\,(1+b\,R^2)^{2\,n}]}{(1+b\,R^2)^{-2\,n}\,[1+a\,R^2\,(1+b\,R^2)^{2\,n}]}\right].
\end{equation}
We should also note that in the relativistic limit the mass-radius ratio
is an important source of information and classification criterion for compact objects \cite{Buchdahl1959}.

\subsection{Energy conditions}
Based on the present work, let us examine the energy conditions
within the framework of general relativity (GR) in the interior of the star. Now, considering the
usual definition of this energy condition for anisotropic fluids, we examine:
(i) the null energy condition (NEC); (ii) the weak energy condition (WEC); and (iii)
the strong energy condition (SEC), at all points in an interior of a star. More precisely, we have the following proposition:
\begin{eqnarray}
\textbf{NEC:}~ \rho+\frac{E^2}{8\pi}\geq 0,\label{eq22}\\
\textbf{WEC}_r:~ \rho+p_r \geq  0 \label{eq23},\\
\textbf{WEC}_t:~ \rho+p_t  +\frac{E^2}{8\pi} \geq 0,\label{eq231}  \\
\textbf{SEC:}~ \rho+p_r+2\,p_t+\frac{E^2}{4\pi} \geq  0.\label{eq24}
\end{eqnarray}

\end{multicols}

\begin{figure}[h]
\centering
\includegraphics[width=8cm]{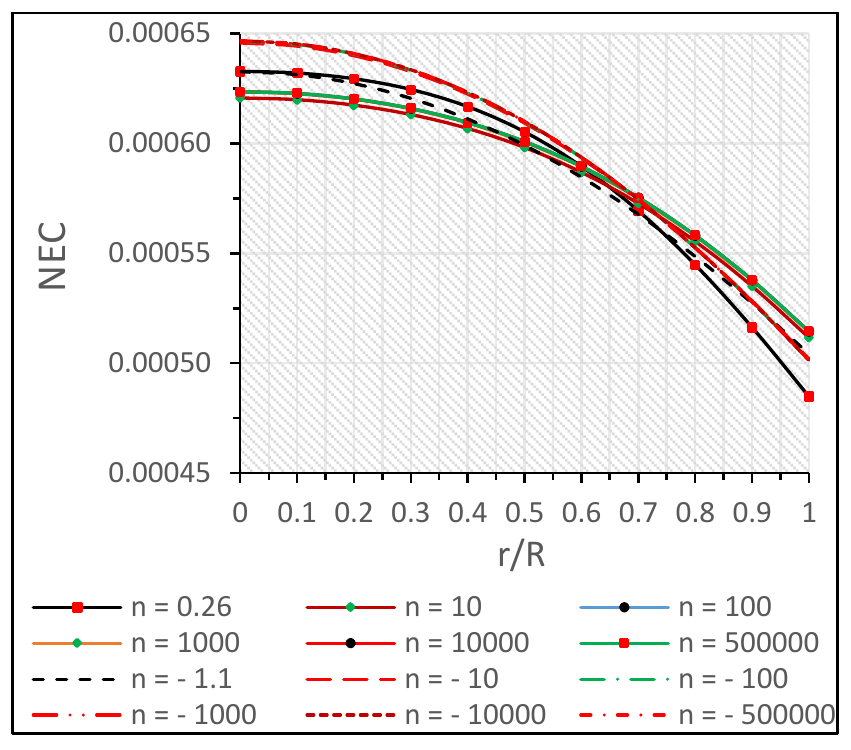} \includegraphics[width=8cm]{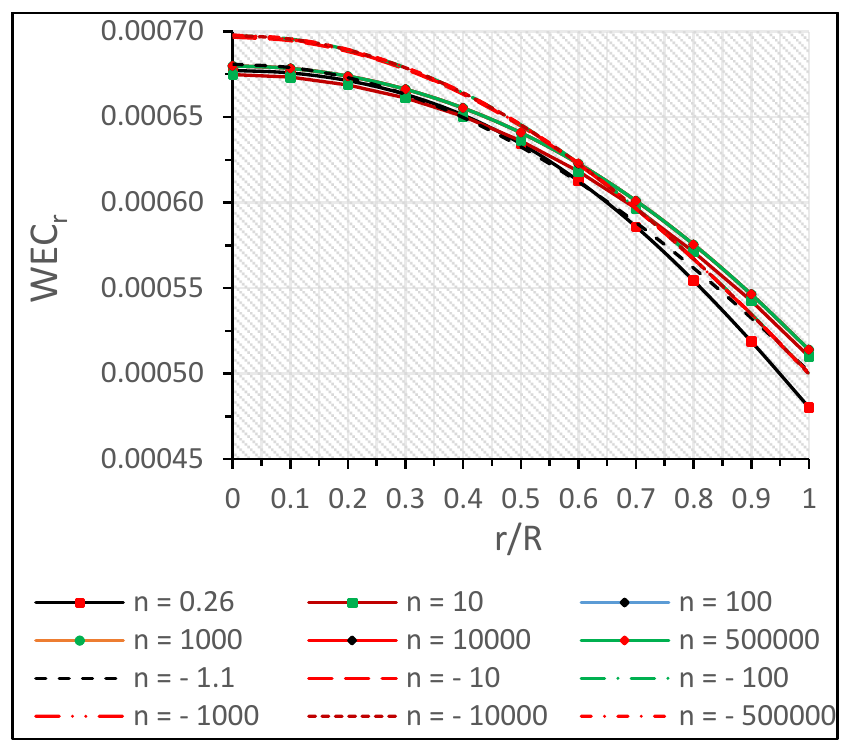}\\
\includegraphics[width=8cm]{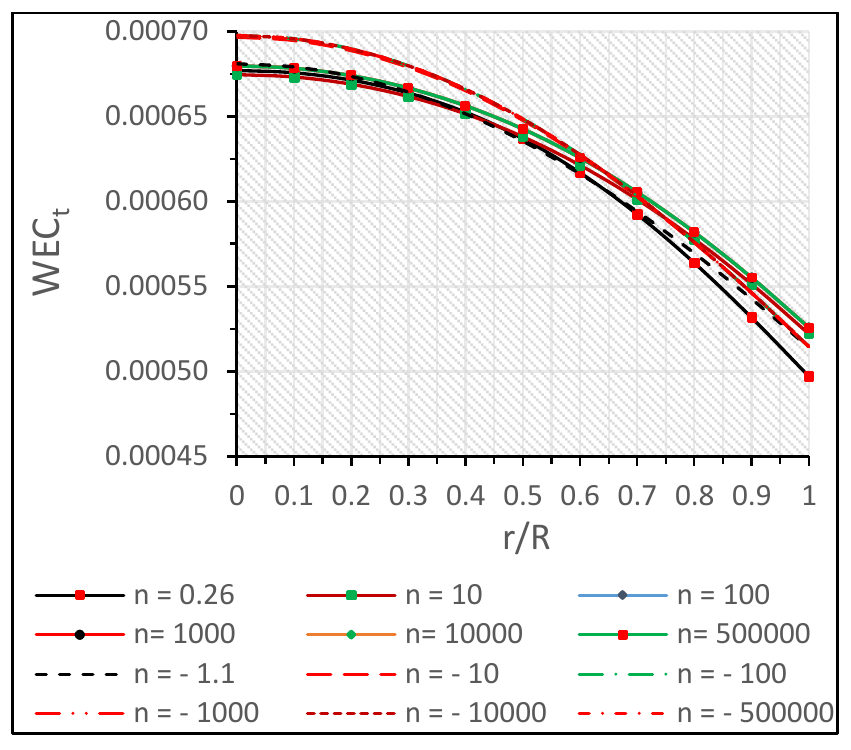} \includegraphics[width=8cm]{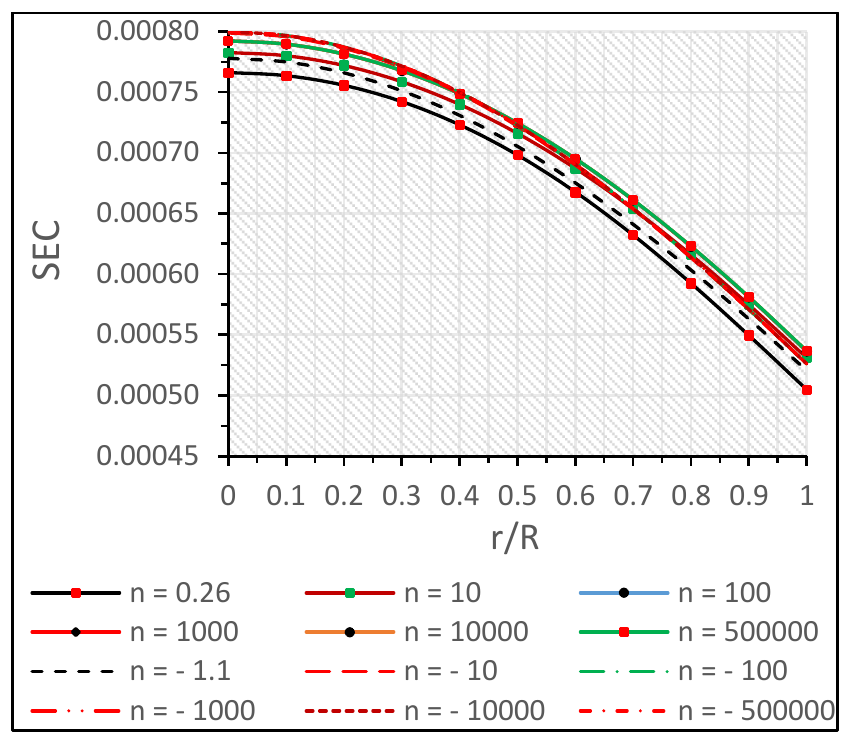}
\caption{Behavior of energy conditions verses fractional radius $r/R$ for Her X-1. The numerical values of the constants are given in Tables 1 and 2.} \label{f3}
\end{figure}

\begin{multicols}{2}
Using the above expressions for all terms in this inequality, one can easily
justify the nature of energy condition for the specific stellar configuration Her X-1.
Graphically, the behaviors of energy conditions are shown in Fig.~3.
For the expressions given in Eqs.~(\ref{eq22}-\ref{eq24}), we only write down the
inequalities and plot the energy conditions as a function of the radius.
As a result, it is eminently clear from Fig.~3 that all energy conditions are satisfied for our proposed model.

\subsection{Equilibrium condition}
We will now proceed further by investigating hydrostatic
equilibrium for different forces acting on the star. These are gravitational, hydrostatic, anisotropic
and electric forces. Hence in the present approach we focus on the generalized
Tolman-Oppenheimer-Volkoff (TOV) equation \cite{Tolman1939,Oppenheimer1939}, 
given by
\begin{equation}
-\frac{M_G(\rho+p_r)}{r^2}e^{\frac{\lambda-\nu}{2}}-\frac{dp}{dr}+
\sigma e^{\frac{\lambda}{2}}\frac{q}{r^2}+\frac{2\,\Delta}{r} =0, \label{eq25}
\end{equation}
where the effective gravitational mass $M_G(r)$ is defined by
\begin{equation}
M_G(r)=\frac{1}{2}r^2 \nu^{\prime}e^{(\nu - \lambda)/2}.\label{eq26}
\end{equation}
Now, using the expression for $M_G(r)$ in Eq.~(\ref{eq25}), we have
\begin{equation}
-\frac{\nu'}{2}(\rho+p_r)-\frac{dp}{dr}+\sigma \frac{q}{r^2}e^{\frac{\lambda}{2}} +\frac{2\,\Delta}{r}=0.  \label{eq27}
\end{equation}
Interestingly, Eq.~(37) may be expressed as a sum of
different components of forces, namely,  gravitational forces $\left(F_g = -\nu'(\rho+p_r)/2\right)$,
hydrostatic $(F_h=-dp_r/dr)$, electric $(F_e=\sigma\,q\,e^{\frac{\lambda}{2}}/r^2)$ and anisotropic $(F_a=2\,\Delta/r)$ forces, respectively. Plugging in the typical values from Eqs. (23-25), and using the
above expressions, we get the different forces in a straightforward way, which leads to:
\end{multicols}

\begin{eqnarray}
F_g(r) &=& -\frac{b\,r\,(n+1)\,B\,\Psi^{2n}}{\pi\,[A+B\,\Psi^{n+1}]}\,\left[\frac{ a\,B\,\Psi^{2n}(1+3\,b\,x+ 4\,n\,b\,x)+F_{g1}(x)]}{2\,[1+a\,x\,\Psi^{2n}]^2\, (A + B\,\Psi^{n+1})} \right], \nonumber \\
F_h(r)&=& \frac{r}{\pi}\left[\frac{ [ B\,F_{h1}(x) + A\,F_{h2}(x)]}{\,(1 + a\, x\,\Psi^{2 n})^2 \,(A + B\, \Psi^{1 + n})^2} -\frac{a\,\,F_{h3}(x)}{4(1+a\,x\,\Psi^{2 n})^2}-\frac{q_0\,a^2 x (\Psi + n\,b\,x)}{2\,\Psi^{1-2n}}\right], \nonumber \\
F_e (r)&=& \frac{q_0\,a^2\,r^3\,\Psi^n\,[\,2\,\Psi^n+n\,b\,x\,\Psi^{n-1}]}{2\,\pi}, \nonumber \\
F_a(r)&=& \frac{r}{4\,\pi}\,\left[\frac{-q_0\,a^2\,x\,\Psi^{2\,n}}{2}+ \frac{\Psi^{n-1}\, (-2\, b\, n + a\,\Psi^{2 n} +
   a\,b\,x\,\Psi^{2 n})\,F_{a1}(x)}{4\,(1 + a\, x\,\Psi^{2 n})^2\,(A + B\, \Psi^{1 + n})}\,\right],
\end{eqnarray}

where we have defined
\begin{eqnarray}
F_{g1}(x) = 2\,b\,B\,(n+1)+a\,A\,\Psi^{n-1}\,(1+b\,x +2\,n\,b\, x),\\
F_{h1}(x)= b\,B (n+1)\,\Psi^{2n}\,[a\,\Psi^{2 n}+b\,(1 + 2\,a\,(n+1)\,x\,\Psi^{2 n})],\\
F_{h2}(x) = b\,B (n+1)\,\Psi^{n-1}\,[ a\,\Psi^{2 n}+ b\,(a\,x\,\Psi^{2 n} + n (-1 + a\,x\,\Psi^{2 n}))],\\
F_{h3}(x)=\Psi^{2 n-1}\,(-2\,b\,n + a\,\Psi^{2 n} + a\,b\,x\,\Psi^{2\,n}),\\
F_{a1}(x)=[a\,\Psi^n\,(A + B\,\Psi^n)-b\,B\,(2 + 2\,n - a\, x\,\Psi^{2 n})].
\end{eqnarray}

\begin{figure}[h!]
\centering
\includegraphics[width=5cm]{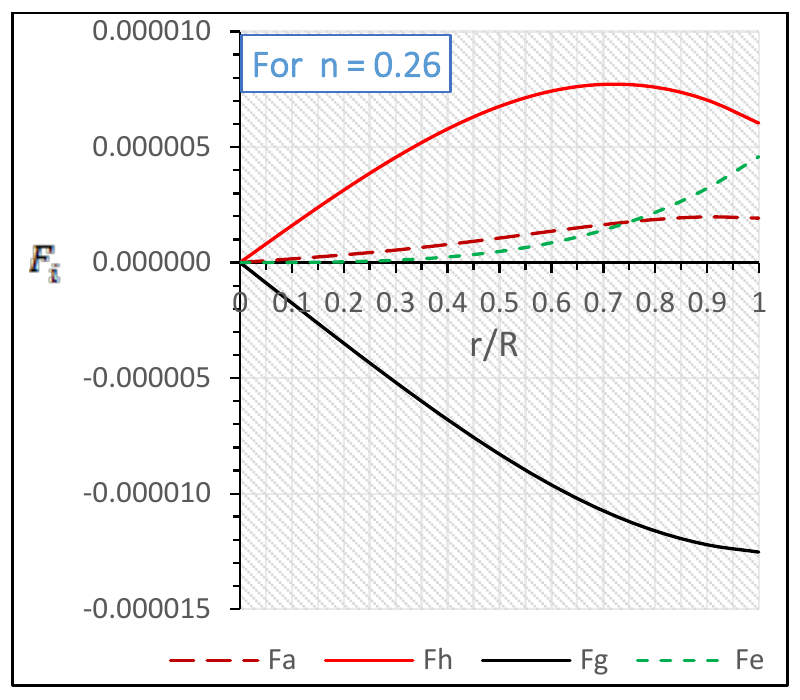}
\includegraphics[width=5cm]{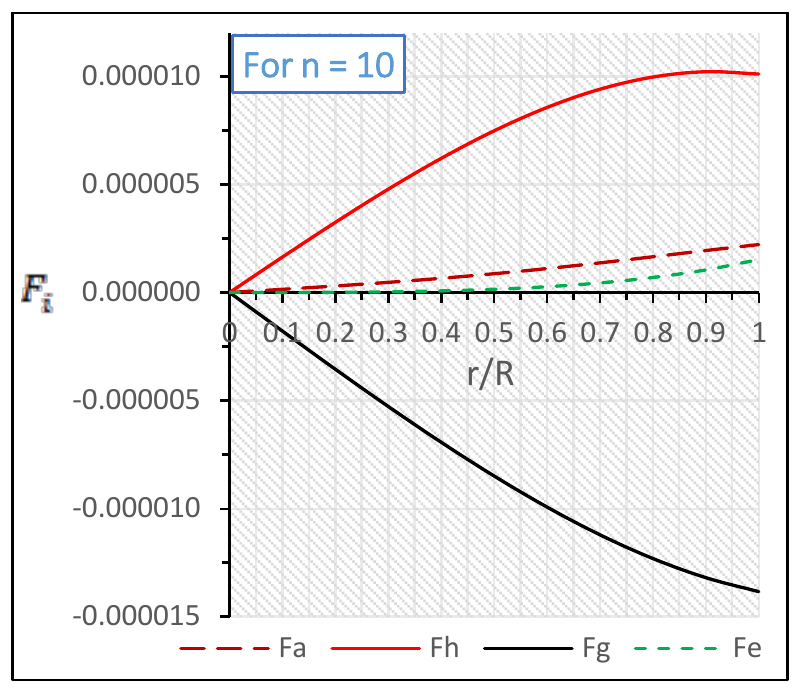}
\includegraphics[width=5cm]{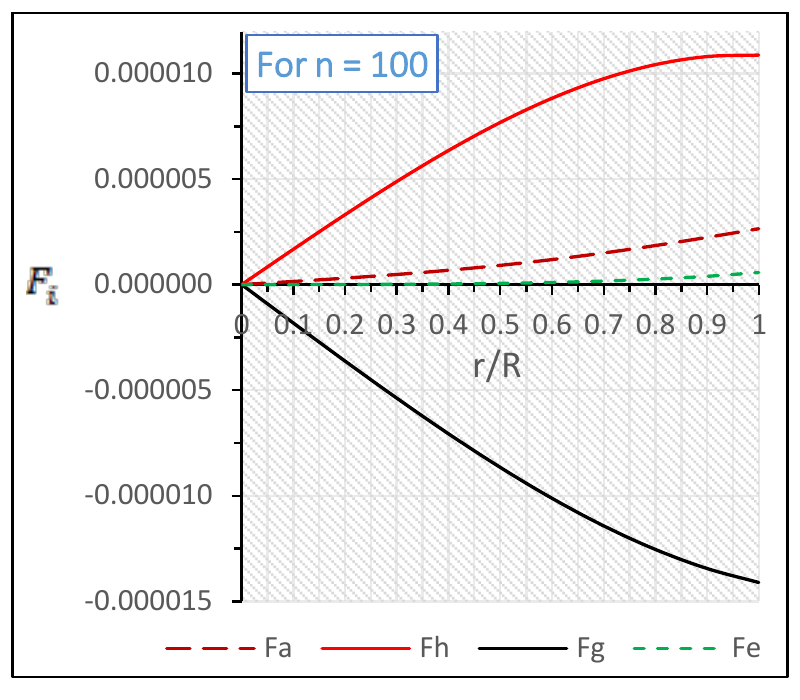}
\includegraphics[width=5cm]{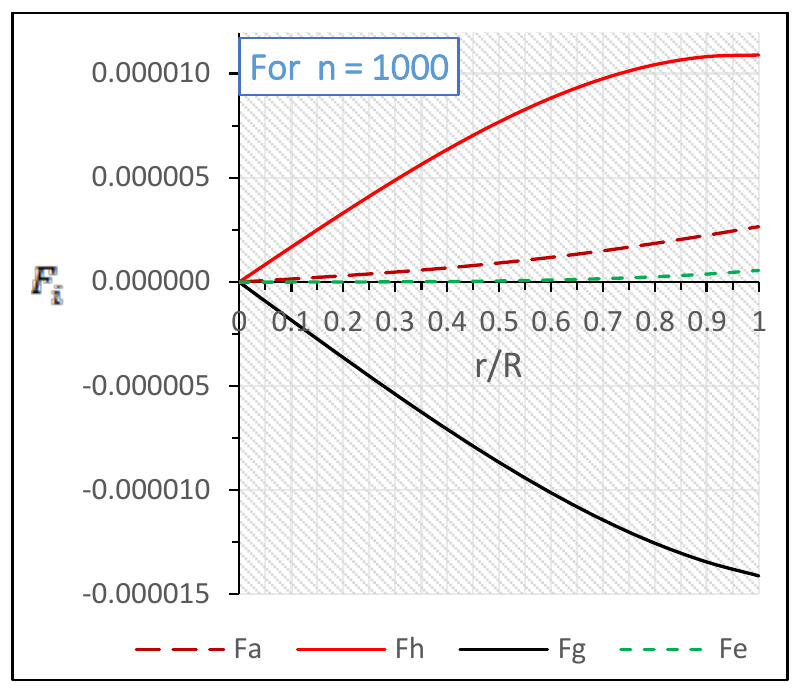}
\includegraphics[width=5cm]{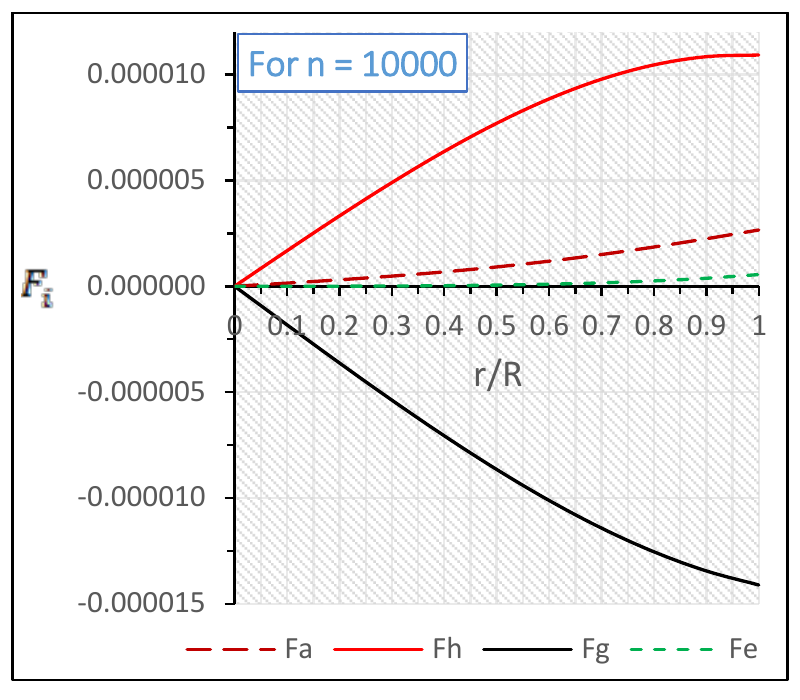}
\includegraphics[width=5cm]{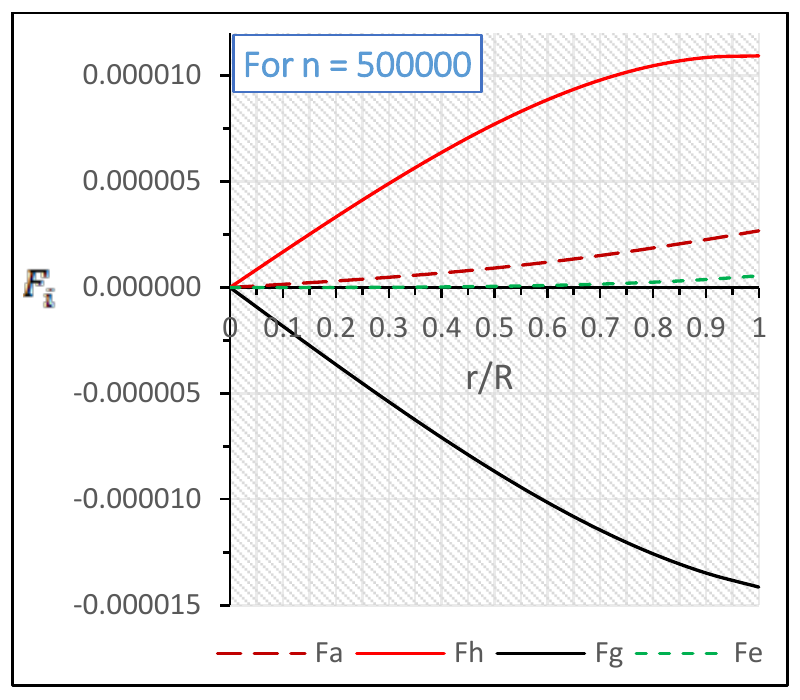}
\includegraphics[width=5cm]{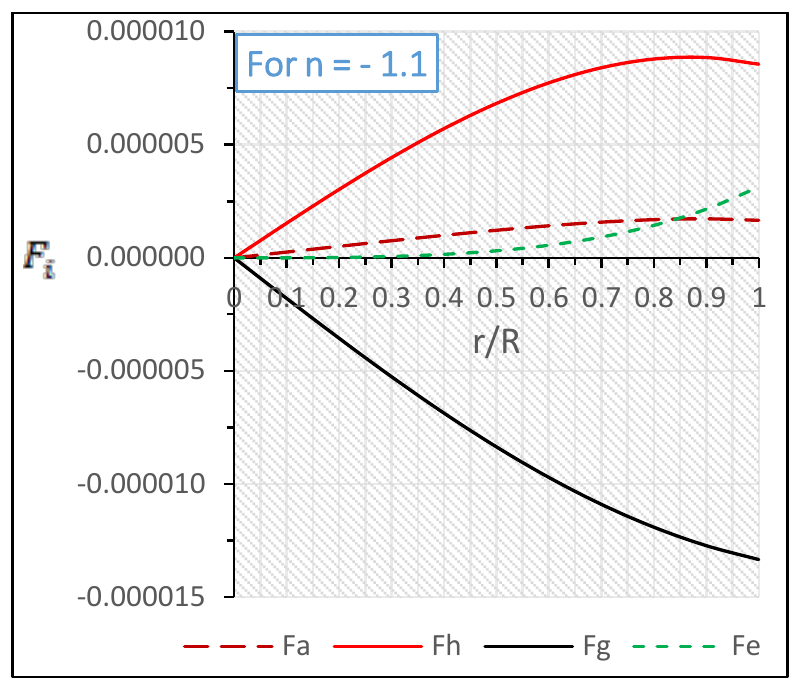}
\includegraphics[width=5cm]{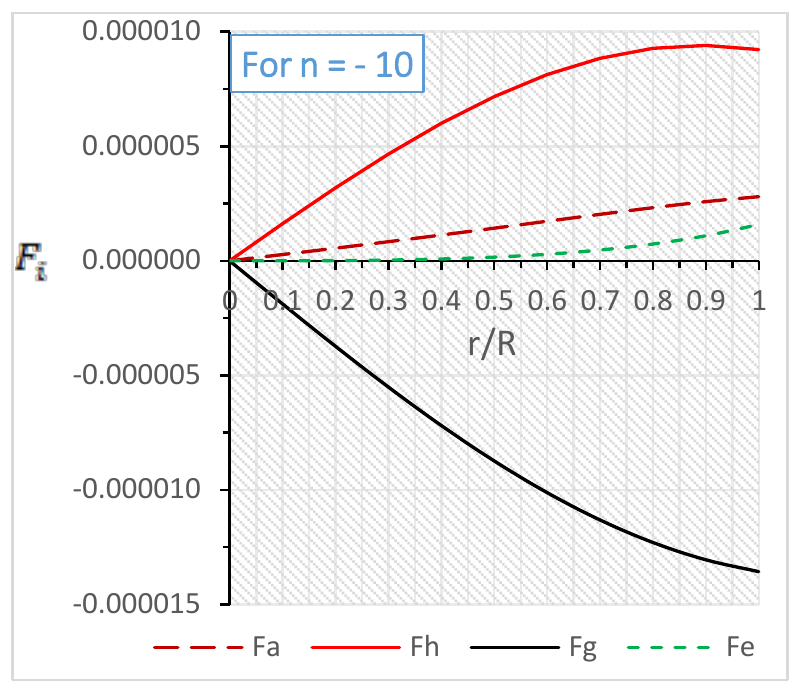}
\includegraphics[width=5cm]{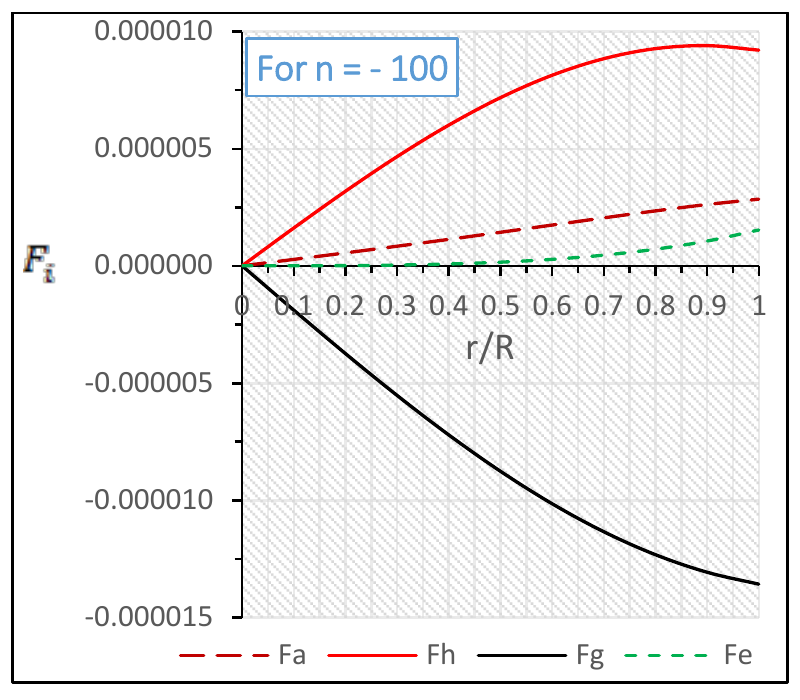}
\includegraphics[width=5cm]{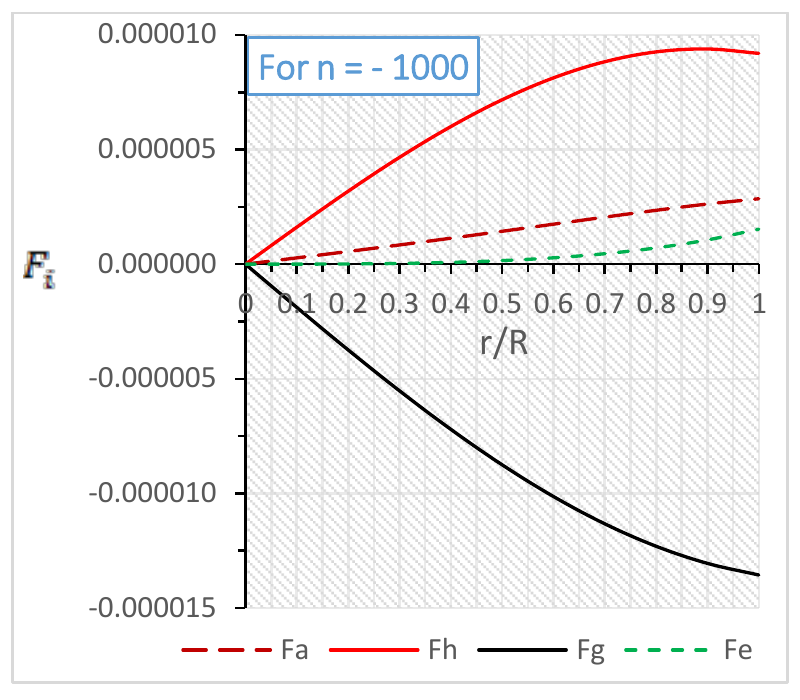}
\includegraphics[width=5cm]{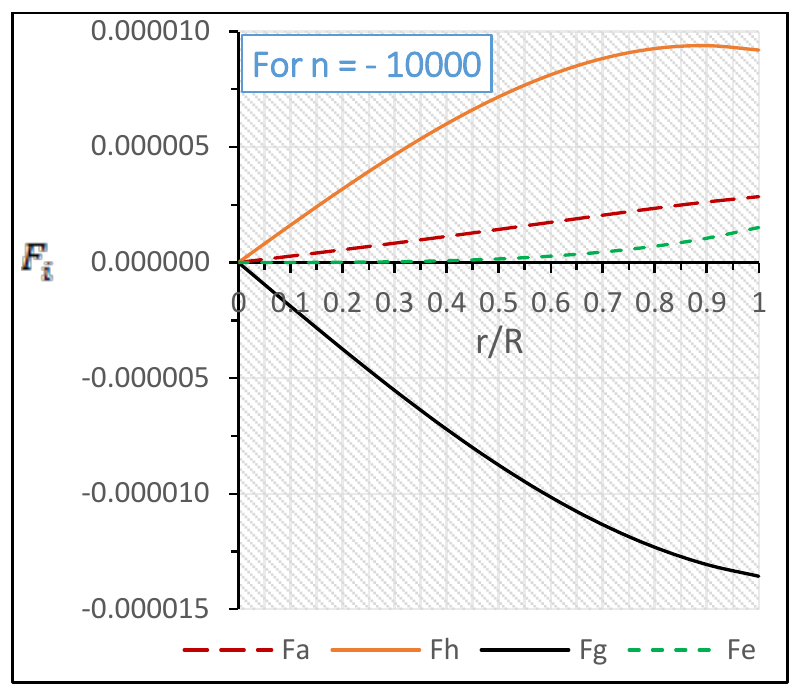}
\includegraphics[width=5cm]{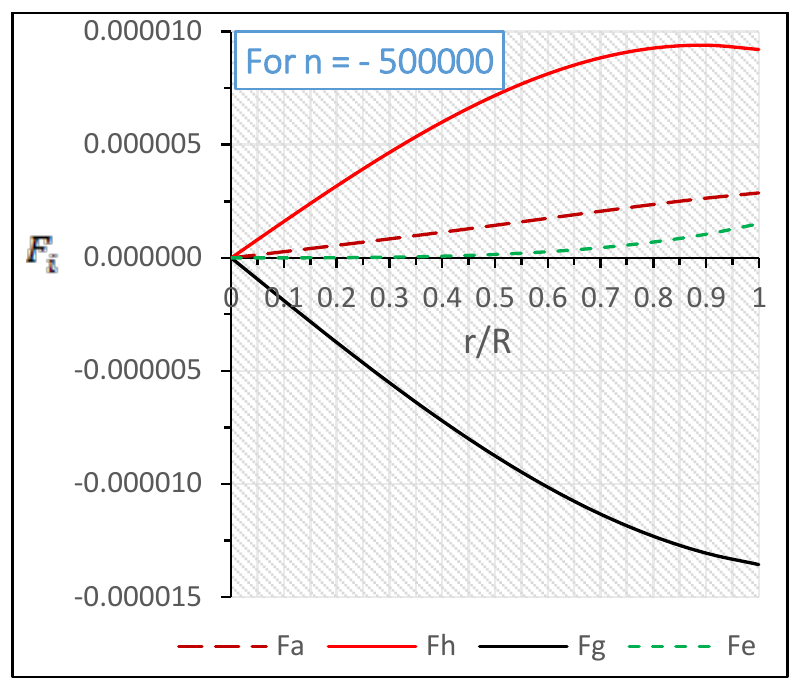}
\caption{Behavior of different forces $F_i$ verses fractional radius $r/R$ for Her X-1. The numerical values of the constants are given in Tables 1 and 2.} \label{f4}
\end{figure}

\begin{multicols}{2}
It is worth noting that when combining all results, we obtain
\begin{equation}
F_g+F_h+F_e+F_a=0. \label{eq28}
\end{equation}
Due to the structure of the obtained modified TOV equation,
the configuration is in static equilibrium under four different forces. As we see from
Fig.\,\ref{f4}, all four forces maintain an overall equilibrium situation for
positive as well as negative values of $n$. Note that the electric force seems to be
negligible in this balancing mechanism.

\subsection{Stability analysis}
Now, we consider $c^2 = dp/d\rho$, which can be  interpreted as the speed
of sound propagation. The motivation behind such a construction is to study the
stability of the configuration. Note that $c^2<0$ is usually
interpreted as an instability, while $c^2 > 1$ indicates superluminal speed of sound.
Our aim here is to verify that the speed of sound  does not exceed the speed of light, i.e., $0 \leq c^2< 1$.
Here, we shall investigate the speed of sound for charged anisotropic fluid distribution satisfying the bounds $0 < v_{r}^{2}={dp_r}/{d\rho} < 1$ and $0 < v_{t}^{2}={dp_t}/{d\rho} < 1$, as in Ref. \cite{Herrera(2016)}.
The velocity of sound in the radial and transverse directions is  given respectively by:

\end{multicols}

\begin{eqnarray}
\hspace{-1.4 cm}
v^2_r &=& \frac{f_1(x)}{f^2_2(x)}\,\left[\frac{B\,v_1(x)+ A\,v_2(x)+a\,\,v_3(x)\,f^2_2(x)-2\,q_0\,a^2\,f^2_2(x)\,v_4(x)}{\,a\,\Psi^{2n-2}
[\,a\,v_5(x)+ b^2\,x\,v_6(x)\,]+2\,q_0\,a^2\,v_4(x)}\right], \nonumber \\
v^2_t &=&\frac{\Psi^{n-2}}{f^2_2(x)}\,\left[\frac{v_8(x)+v_9(x)+v_{10}(x)+v_{11}(x)+ b\,f_1(x)\,f_2(x)\,v_{12}(x)+v_{13}(x)}{-\,a\,\Psi^{2n-2}
[\,a\,v_5(x)+ b^2\,x\,v_6(x)\,]+2\,q_0\,a^2\,v_4(x)}\right].
\end{eqnarray}
where the unknown quantities stand for
\begin{eqnarray*}
f_1(x) &=& (1 +a\,x\,\Psi^{2 n}),  \,\, f_2(x)=(A +B\,\Psi^{n+1}), \nonumber \\
v_1(x)&=& 4\,b\,B\,(n+1)\,\Psi^{2n}\,[a\,\Psi^{2 n}+b\,(1 + 2\,a\,(n+1)\,x\,f_1(x))], \nonumber \\
v_2(x)&=& 4\,b\,B\,(n+1)\,\Psi^{n-1}\,[a\,\Psi^{2 n}+b\,(a\,x\,\psi^{2 n} + n\,(-1 + a\,x\,\Psi^{2 n})), \nonumber \\
v_3(x)&=& \Psi^{2 n-1}\,(-2\,b\,n + a\,\Psi^{2 n} + a\,b\,x\,\Psi^{2 n}), v_4(x)=x\,\Psi^{2 n-1}\,(\Psi + b\,n\,x)\,f^2_1(x), \nonumber \\
v_5(x)&=& \Psi^{2 n}\,(5 + a\,x\,\Psi^{2 n})+ 2\,b\,[a x \Psi^{2 n}
(5 + a\,x\,\Psi^{2 n})+n\,(-5 + 3\,a\,x\,\Psi^{2 n})],\nonumber \\
v_6(x)&=& [ 8 n^2 (-1 + a\,x\,\Psi^{2 n}) + a\,x\,\Psi^{2 n}\,(5 + a\,x\,\Psi^{2 n}) + 2\,n\,(-3 + 5\,a\,x\,\Psi^{2 n})],\nonumber \\
v_7(x)&=& b\,(4\,B\,(n+1)-a\,A\,(1+2\,n)\,x\,\Psi^{n}+b^2\, B\,x\,(4 + 8\,n + 4\,n^2\,+ a\,x\,\Psi^{2 n}),\nonumber \\
v_8(x)&=& -b\,B\, (n+1)\,\Psi^{n+1}\,f_1(x)\,[-a \Psi^{^n}\,(A + B \Psi^{n})+v_7(x)], \nonumber \\
v_9(x)&=& b\,(n-1)\,f_1(x)\,f_2(x)\, [-a\,\psi^n (A + B\,\Psi^n) +v_7(x)], \nonumber \\
v_{10}(x)&=& -2\,a\,\Psi^{2 n}\,f_2(x)\,(\Psi + 2\,b\,n\,x)\,[-a\,\Psi^n\,(A + B\,\Psi^n) +v_7(x)], \nonumber \\
v_{11}(x)&=&  b\,[-a\,A\,(1 + 3 \,n + 2\, n^2)\,x \,\Psi^n + 2\,B\,(2 + 4\,n + 2\,n^2 + a\,x\,\Psi^{2 n})], \nonumber \\
v_{12}(x)&=& [-a\,\Psi^n (A + 3\,A\,n + 2\,B\,n\,\Psi^n)+2\,b^2\,B\,(1 + n)\,x\,(2 + 2\,n + a\,x\,\Psi^{2 n}) + v11(x)],\nonumber \\
v_{13}(x)&=& 2\,q_0\,x\,a^2\,f^3_1(x)\,f^2_2(x)\,\Psi^{n+1}\,(\Psi + b\,n\,x).
\end{eqnarray*}

\begin{figure}[h]
\centering
\includegraphics[width=8cm]{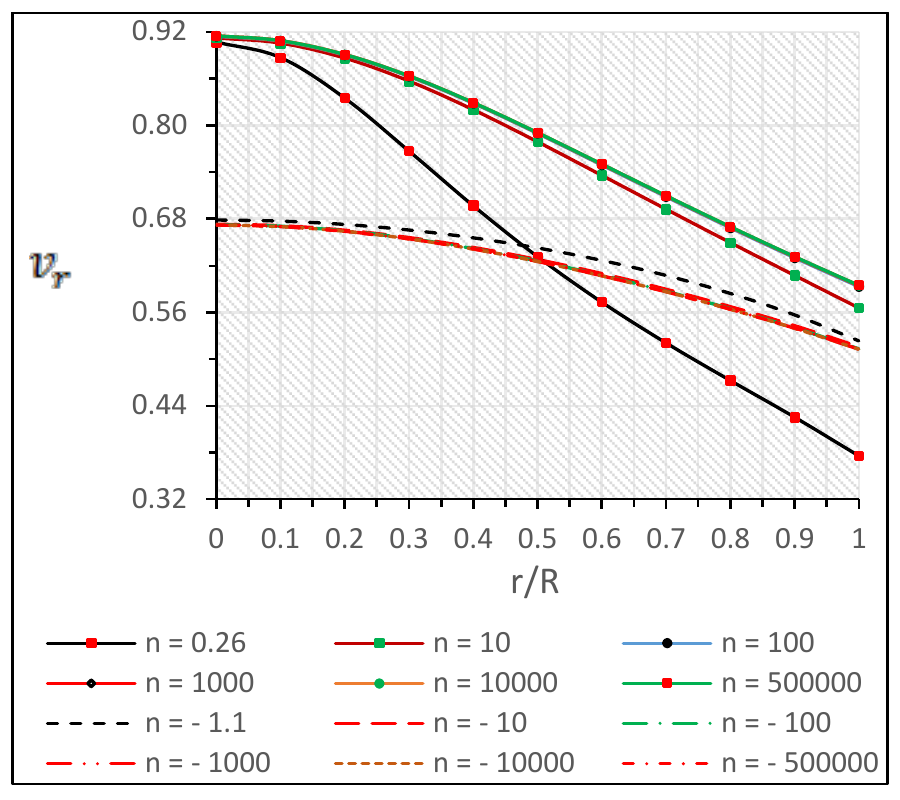} \includegraphics[width=8cm]{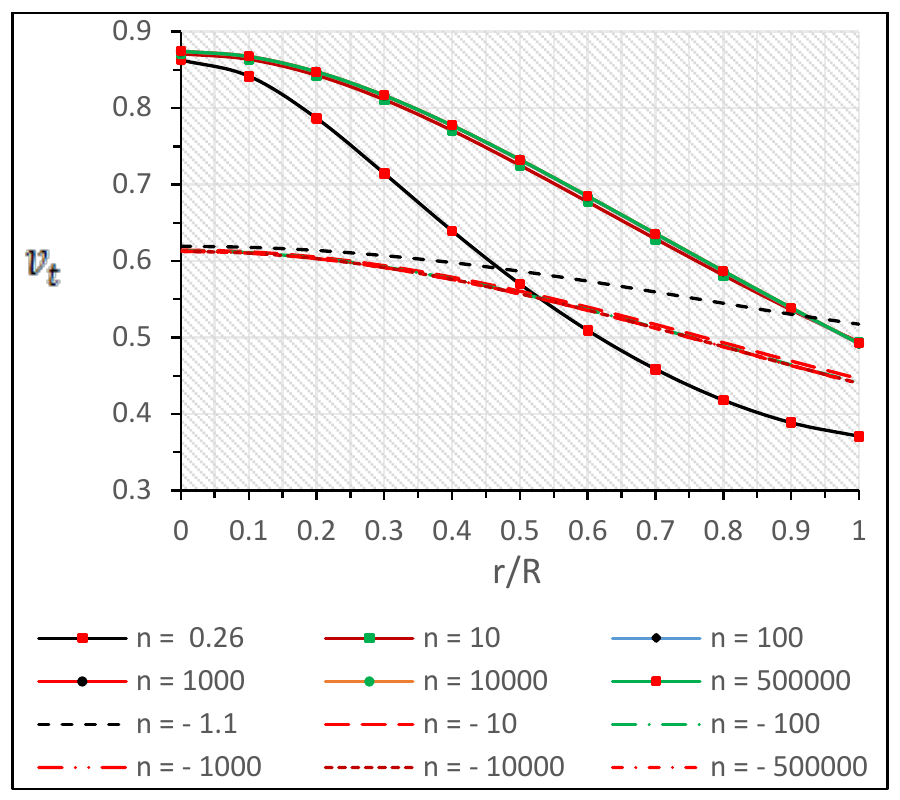}
\caption{Behavior of radial velocity ($v_r$) and tangential velocity ($v_t$) verses fractional radius $r/R$ for Her X-1. The numerical values of the constants are given in Tables 1 and 2.} \label{f5}
\end{figure}

\begin{multicols}{2}

Our results are displayed in Fig.\,\ref{f5}. Looking at the radial/transverse pressures versus radius, it
is interesting to note that both $v_r^2,\,v_t^2 < 1$ and decrease monotonically
from the centre of the star  towards the surface of the star. Moreover, both velocities
are positive and finite at the interior of the star Her X-1.

\subsection{Electric charge}
\begin{figure}[h]
\centering
\includegraphics[width=8cm]{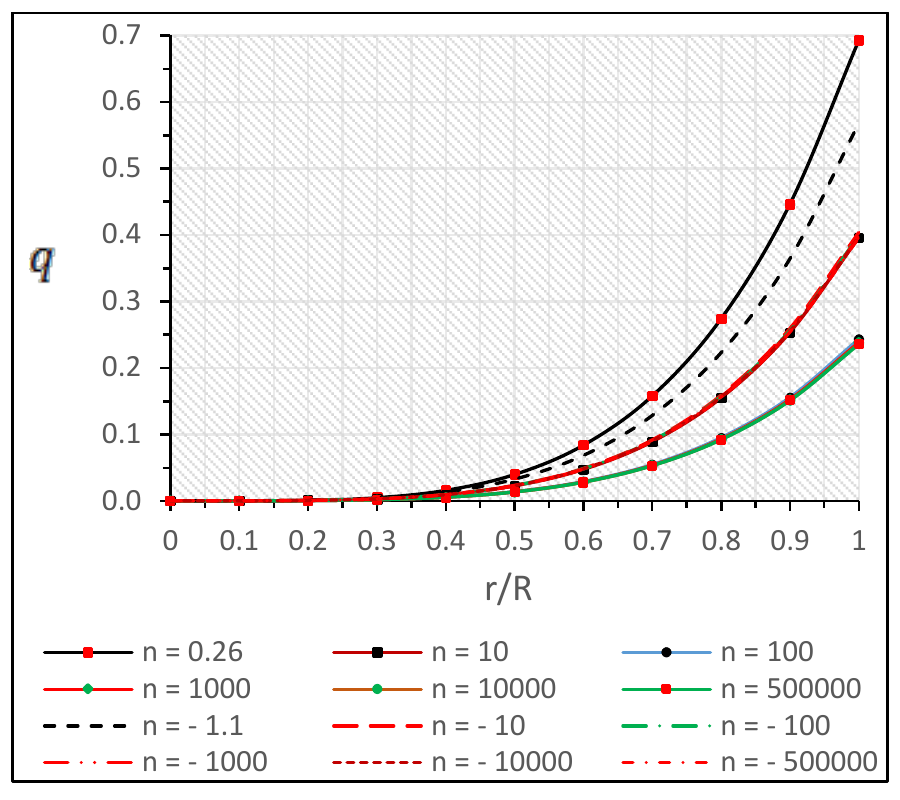}\includegraphics[width=8cm]{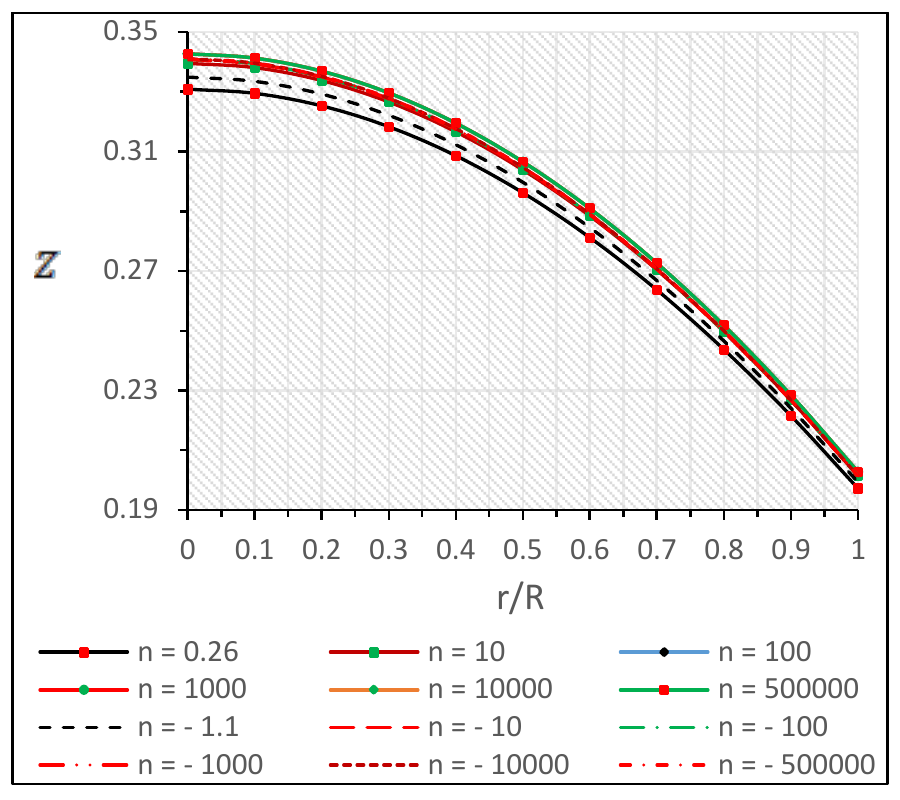}
\caption{(Left) Behavior of electric charge ($q$) versus fractional radius $r/R$ for Her X-1. (Right) Behavior of redshift ($Z$) verses fractional radius $r/R$ for Her X-1. The numerical values of the constants are given in Tables 1 and 2.} \label{f6}
\end{figure}

Our aim here is to explore how the physics governing the structure of stars, includes the charge, affects the structure and stability of strange stars. For this purpose we use the graphical representation given in Fig. \ref{f6}. It is clear from Fig. \ref{f6}, and Tables \ref{Table1} and \ref{Table2}, that electric charge is zero at the centre i.e. E(0)=0, which corresponds to vanishing of the electric field at the center. Moreover, we observe that electric charge distributions are monotonically increasing away from the center, and reach a maximum value at the surface. Depending on the distribution of charge, the inner core is less stable than the surface of the charged body. It is important to note that charge  decreases monotonically with an increase of $|n|$, with the difference becoming indistinguishable at the stellar surface for very large values of $|n|$. According to Ref. \cite{M6}, we can interpret $n$ as a \lq\lq stabilizing factor\rq\rq. It can be found by searching for the condition that higher charge configurations reach for large radius and large values of the metric coefficient $e^{\lambda}$. From this we understand that the matter and charge are related to each other and the relations are strongly coupled. As mentioned in Ref. \cite{M6}, the inclusion of anisotropic pressure into the matter also enables us to find exact electric star solutions that are held against collapse by electric repulsion, and avoid black hole formation.

The charged compact stars that appear in this study have a TOV  limit, and the electric and gravitational forces are balanced. According to Ref. \cite{rayb}, the net charge can be as high as $10^{20}$ C, to see any considerable effect on the phenomenology of compact stars.  Such objects live within the very short period of time between a supernova explosion and the formation of a charged black hole. However, Bekenstein \cite{bek} showed that the effect of charge distribution can be seen in the general relativistic hydrostatic equilibrium. If this is the case, pair production may be induced within the star and thus destabilize the core. To be more concrete, we calculate the amount of charge at the boundary in units of coulombs for the compact star Her X-1 as follows: (i-1) $8.0760 \times10^{19}$ C for $n=0.26$, (ii-1) $4.6186 \times 10^{19}$ C for $n=10$, (iii-1) $2.8391 \times 10^{19}$ C for $n=100$, (iv-1) $2.7752 \times 10^{19}$ C for $n=1000$, (v-1) $2.7592 \times 10^{19}$ C for $n=10000$, (vi-1)  $2.7503 \times 10^{19}$ C for $n = 500000$ and (i-2) $6.6211 \times 10^{19}$ C for $n=-1.1$, (ii-2) $0.4710 \times 10^{19}$ C for $n=-10$, (iii-2) $4.6450 \times 10^{19}$ C for $n=-100$, (iv-2) $4.6373\times10^{19}$ C for $n=-1000$, (v-2) $4.6238 \times 10^{19}$ C for $n=-10000$, and (vi-2)  $4.6104 \times 10^{19}$ C for $n=-500000$. For different values of n, the charge is shown in Tables \ref{Table1} and \ref{Table2}, determined by multiplying every listed value by a factor of $1.1659 \times 10^{20}$.

\end{multicols}

\begin{table}
\centering \caption{The electric charge (in coulombs) for compact star Her X-1 for different positive values of $n$.} \label{Table1}

{\begin{tabular}{@{}ccccccc@{}} \hline

 $r/a$  &  $n$ = 0.26  & $n$ =  10  &  $n$ = 100  &  $n$ = 1000  & $n$ = 10000  &  $n$ = 50000  \\ \hline

0.2 &0.0009985   & 0.0005589 & 0.0003432 & 0.0003354 & 0.0003335 & 0.0003324  \\ \hline

0.4 &0.0162269   & 0.0090854 &  0.0055789 & 0.0054528 & 0.0054213 & 0.0054038  \\ \hline

0.6 &0.0841471   & 0.0472200 &  0.0290002   & 0.0283450  & 0.0281812 &0.0280900  \\ \hline

0.8 &0.2741287   & 0.1548149 & 0.0951107   & 0.0929640 & 0.0924270 & 0.0921279  \\ \hline

1.0 &0.6926801    & 0.3961417 & 0.2435158  & 0.2380303 & 0.2366565 & 0.2358908  \\ \hline

\end{tabular}}
\end{table}

\begin{table}
\centering \caption{The electric charge (in coulombs) for compact star Her X-1 for different negative values of $n$.} \label{Table2}

{\begin{tabular}{@{}ccccccc@{}} \hline

 $r/a$  &  $n$ = -1.1  &  $n$ =  -10   &   $n$ = -100  &   $n$ = -1000  & $n$ = -10000  &  $n$ = -50000  \\ \hline

0.2 &0.0008130   & 0.00058101 & 0.0005733 & 0.0005724 & 0.0005707 & 0.0005691  \\ \hline

0.4 &0.0131827   & 0.009420316 & 0.0092952 & 0.0092802 & 0.0092533 & 0.0092265  \\ \hline

0.6 &0.0682580    & 0.0487584 &  0.0481082 & 0.0480300 & 0.0478910 & 0.0477524  \\ \hline

0.8 &0.2228174    & 0.1589667 & 0.1568228  & 0.1565654 & 0.1561122 & 0.1556601  \\ \hline

1.0 &0.5678978    & 0.4039877 & 0.3984088  & 0.3977420 & 0.3965893 & 0.3954407  \\ \hline

\end{tabular}}
\end{table}

\begin{multicols}{2}

\subsection{Effective mass and compactness parameter for a charged compact star}
Now we begin with the mass-radius ratio $M/R$ of a relativistic compact star.
The bounds on a spherically symmetric isotropic fluid sphere are given
by $2\mathrm{M/R}\leq 8/9 $, according to Buchdahl \cite{Buchdahl1959}. We note that for a
compact charged fluid sphere there is a  lower bound for the mass-radius ratio, which is \cite{Boehmer2006}
\begin{equation}
\frac{Q^{2}\, (18 R^2+ Q^2) }{2R^{2}\, (12R^2+Q^2)}  \leq
\frac{M}{R}, \label{eq62}
\end{equation}
for $Q < M$. This result was generalized in Ref.~\cite{And}, for new
bounds on the mass-radius ratio for charged compact objects, with the following relation
\begin{equation}
\frac{M}{R} \leq \left[\frac{4R^2+3Q^2}{9R^2} +\frac{2}{9R}\,\sqrt{R^2+3Q^2}\right]. \label{eq63}
\end{equation}
Now, using  Eqs.~(\ref{eq62}) and (\ref{eq63}), we obtain the mass-radius ratio
within the region
\begin{equation}
\frac{Q^{2}\, (18 R^2+ Q^2) }{2R^{2}\, (12R^2+Q^2)}  \leq
\frac{M}{R} \leq \left[\frac{4R^2+3Q^2}{9R^2} +\frac{2}{9R}\,\sqrt{R^2+3Q^2}\right].
\end{equation}
To obtain the effective mass for a charged fluid sphere within the radius $r=$R, one can then compute
\begin{equation}
m_{eff}=4\pi{\int^R_0{\left(\rho+\frac{E^2}{8\,\pi}\right)\,r^2\,dr}}=\frac{R}{2}[1-e^{-\lambda(R)}]\, ,\label{eq32}
\end{equation}
where $e^{-\lambda}$ is given by Eq.~(\ref{e4}). Using the
above results, the compactness factor can be written as
\begin{equation}
u(R)=\frac{m_{eff}(R)}{R}=\frac{1}{2}[1-e^{-\lambda(R)}]\, .\label{eq33}
\end{equation}
Let us now consider the surface redshift, which allows us to write down
the following expression
\begin{equation}
Z_s= (1-2\,u)^{\frac{-1}{2}} -1=\sqrt{1+D\,AR^2\,(1-AR^2)}-1. \label{zs}
\end{equation}
Note that for the isotropic stellar configuration the maximum possible surface redshift
is $Z_s$ = 4.77, but it may be exceeded in the presence of  anisotropy pressure according to Ref.~\cite{bowers}. In our model, this condition falls within the limit for isotropic pressure given in
Tables \ref{Table2} and \ref{Table11-1}. The above study is carried out for both
positive and negative values of $n$ shown in Fig.\,\ref{f6}.
Table \ref{Table11-1} gives the solutions of the numerical treatment used in this study that allow the existence of a
physically viable model with ansatz for the metric potential (\ref{e4}) for the compact star Her X-1, within a given range of parameters.

\end{multicols}

\begin{table}
\centering \caption{List of solutions with a given range of parameters that give a physically viable model with ansatz for the metric potential (19) for compact star Her X-1.} \label{Table11-1}
{\begin{tabular}{@{}ccccc@{}} \hline

  Range of  &  Electric charge   & Anisotropy Pressure  &  Velocity of Sound  & List of   \\
   $ n  \& b $ &  ($E $ )  &  ($\Delta$ )   & ($dp_i/d\rho < 1$ ) &  Reference \\ \hline

n, b $\in \Re^{+} \cup 0 ~~$  &  E=0  & $\Delta \ne 0$ & Yes~($4\leq n \leq 10$) & \cite{Bhar:2016wlc} \\
$\&$ $~~ n\ne -1$  &     &     &    &   \\ \hline
n, b $\in \Re^{-} \cup 0 $ & $E = 0$ & $\Delta \ne 0$  & Yes~($n \geq 2$) & \cite{SRSS}   \\ \hline

n, b $\in R $  &$E^2=q_0a^2r^4\,(1+br^2)^{2n}$  & $\Delta \ne 0$  & Yes & Present Case  \\
&                             &                      &  $n \in (-\infty, -1.1] \cup [0.26,  \infty)$     & \\  \hline

\end{tabular}}
\end{table}
\begin{multicols}{2}
\section{Discussion}
In this work, we have studied a class of new exact solutions with anisotropic fluid distribution of matter for compact objects in hydrostatic equilibrium. We considered the Einstein- Maxwell system with anisotropic spherically symmetric gravitating source and obtained novel gravitational solutions compatible with observational data for the compact object Her X-1. We matched the interior solution to an exterior Reissner-Nordstr$"o$m solution, and studied some physical features of the models, such as the energy conditions, speed of sound, and mass-radius ratio. We have used the condition arising from embedding a 4-dimensional spherically symmetric static metric in spherical coordinates into a 5-dimensional flat space-time to model a charged object. We employed a so-called class one space-time.

We have found that energy density, radial and  tangential pressures are finite
at the center, and monotonically decreasing functions with respect to the radial co-ordinate for our specific
choice of generic function, which was illustrated in Figs.\ref{f1} and \ref{f2}.
Moreover, the radial pressure vanishes at the boundary of the star, whilst the tangential pressure is non-vanishing at the stellar surface. We notice that the radial and tangential pressures increase with increasing $|n|$
and the energy density is at a maximum at the core of the star. In Tables 1 and 2
we tabulated the calculated values of central and surface density and central pressure for
compact objects considering HER X-1 as a charged anisotropic star.

Our analysis shows that total charge value of a compact object must be around $10^{20}$ C to see any appreciable effect  and the electric fields have to be huge ($\sim$ $10^{20}$) in the equilibrium and stability of the star. More specifically, electric charge that produces significant effect on the structure and stability of the object has its maximum value at the boundary of the star and monotonically increases away from the centre. The electric charge q
 against the radial coordinate has been plotted in Fig.\ref{f6} (left panel) for different parametric values of n.

We further extended our analysis by investigating the energy conditions, hydrostatic equilibrium
under different forces, and velocity of sound. We have found through our analysis that
all energy conditions are satisfied at the interior of the configuration and maintain 
equilibrium between different forces due to anisotropy and electromagnetic effects,
as we can see from  Figs.~\ref{f3} and \ref{f4}. We found that when $|n|$ is small
 the electromagnetic force dominates the anisotropy force near to and at the surface of the star, while when
$|n|$ is large, the anisotropy force dominates the electromagnetic force.

Using the values given in Tables 1 and 2, we also checked the
square of the velocity of sound for anisotropic matter distribution.
As illustrated in Fig.~\ref{f5}, our model obeys causality throughout the stellar interior.
We also pointed out that the surface redshift is higher
in an anisotropic star than an isotropic one, shown on the extreme right of
Tables \ref{Table2} and \ref{Table11-1} for different values of $n$. Furthermore, we showed the maximum allowed
masses with their respective radii i.e., that the mass-radius ratio falls within the limit of
$8/9\,(= 2M/R)$ as proposed by Buchdahl \cite{Buchdahl1959}, for the compact objects considering HER X-1 that we
have considered for our model. Finally, considering charged stars with
specific electromagnetic fields can be used as a tool to probe other compact objects like HER X-1.\\

\textbf{Acknowledgments}: The author S. K. Maurya acknowledges the University of Nizwa for their continuous support and encouragement to carry out this research work.  AB is thankful to the 
Inter-University Centre for Astronomy and Astrophysics, Pune,
India for providing research facilities. We would like to thank the referees for their time and useful comments towards the improvement of our manuscript.\\

{ }

\end{multicols}

\clearpage

\end{document}